\documentclass[journal,10]{IEEEtran}

\usepackage{subfigure}

\usepackage{cite}

\usepackage{graphicx}

\usepackage{multirow}

\usepackage{amsmath, amsfonts, amssymb}
\begin{document}
\title{FiWi Access Networks Based on Next-Generation PON and
  Gigabit-Class WLAN Technologies:\\A Capacity and Delay
  Analysis (Extended Version)\thanks{A shorter version of this paper appears
    in \textit{Proc., IEEE Wireless Communications and Networking
    Conference (WCNC)}, Shanghai, China, April 2013.}\thanks{This work
    was supported by the FQRNT MERIT Short-term Research Scholarship
    Program and NSERC Strategic Project Grant No. 413427-2011.}}

\author{Frank Aurzada\thanks{F. Aurzada is with
FB Mathematik, Technical University Darmstadt, Schlossgartenstr.~7,
64289 Darmstadt, Germany,
      e-mail: aurzada@mathematik.tu-darmstadt.de.}, Martin L\'evesque, Martin
  Maier,~\IEEEmembership{Senior Member,~IEEE},
   \thanks{M. L\'evesque and
    M. Maier are with the Optical Zeitgeist Laboratory, INRS,
    University of Qu\'ebec, Montr\'eal, QC, H5A 1K6, Canada, e-mail:
    \{levesquem, maier\}@emt.inrs.ca.} and Martin
  Reisslein,~\IEEEmembership{Senior Member,~IEEE}
   \thanks{M. Reisslein
    is with Electrical, Computer, and Energy Eng.,
       Arizona State Univ., Tempe, AZ 85287-5706, USA, e-mail:
    reisslein@asu.edu.}}

\maketitle

\begin{abstract}
Current Gigabit-class passive optical networks (PONs) evolve into
next-generation PONs, whereby high-speed 10+ Gb/s time division
multiplexing (TDM) and long-reach wavelength-broadcasting/routing
wavelength division multiplexing (WDM) PONs are
promising near-term candidates. On the other hand, next-generation
wireless local area networks (WLANs) based on frame aggregation
techniques will leverage physical layer enhancements, giving rise
to Gigabit-class very high throughput (VHT) WLANs. In this paper, we
develop an analytical framework for evaluating the capacity and delay
performance of a wide range of routing algorithms in converged
fiber-wireless (FiWi) broadband access networks based on different
next-generation PONs and a Gigabit-class multi-radio multi-channel
WLAN-mesh front-end. Our framework is very flexible and incorporates
arbitrary frame size distributions, traffic matrices, optical/wireless
propagation delays, data rates, and fiber faults. We verify the
accuracy of our probabilistic analysis by means of simulation for the
wireless and wireless-optical-wireless operation modes of various FiWi
network architectures under peer-to-peer, upstream, uniform, and
nonuniform traffic scenarios. The results indicate that our proposed
optimized FiWi routing algorithm (OFRA) outperforms minimum (wireless)
hop and delay routing in terms of throughput for balanced
and unbalanced traffic loads, at the expense of a slightly increased
mean delay at small to medium traffic loads.
\end{abstract}


\section{Introduction}
\label{sec:intro}
\PARstart{F}{iber-wireless} (FiWi) access networks, also referred to
as wireless-optical broadband access networks (WOBANs),
combine the reliability, robustness, and high capacity of optical
fiber networks and the flexibility, ubiquity, and cost savings of
wireless networks~\cite{SYDM09}. To deliver peak data rates up to
200 Mb/s per user and realize the vision of complete fixed-mobile
convergence, it is crucial to replace today's legacy wireline and
microwave backhaul technologies with integrated FiWi broadband access
networks~\cite{AEEH10}.

Significant progress has been made on the design of advanced FiWi
network architectures as well as access techniques and routing
protocols/algorithms over the last few years~\cite{GhMa11a}. Among
others, the beneficial impact of advanced hierarchical frame
aggregation techniques on the end-to-end throughput-delay performance
of an integrated Ethernet passive optical network (EPON)/wireless mesh
network (WMN)-based FiWi network was demonstrated by means of
simulation and experiment for voice, video, and data
traffic~\cite{GhMa11}. A linear programming based routing algorithm
was proposed in~\cite{ZhWW09,ZWW09} with the objective of maximizing
the throughput of a FiWi network based on a cascaded EPON and
single-radio single-channel WMN. Extensive simulations were conducted
to study the throughput gain in FiWi networks under peer-to-peer
traffic among wireless mesh clients and compare the achievable
throughput gain with conventional WMNs without any optical
backhaul. The presented simulation results show that FiWi and
conventional WMN networks achieve the same throughput when all traffic
is destined to the Internet, i.e., no peer-to-peer traffic, since the
interference in the wireless front-end is the major bandwidth
bottleneck. However, with increasing peer-to-peer traffic the
interferences in the wireless mesh front-end increase and the
throughput of WMNs decreases significantly, as opposed to their FiWi
counterpart whose network throughput decreases to a much lesser extent
for increasing peer-to-peer traffic.

The design of routing algorithms for the wireless front-end only or
for both the wireless and optical domains of FiWi access networks has
received a great deal of attention, resulting in a large number
of wireless, integrated optical-wireless, multipath, and energy-aware
routing algorithms. Important examples of wireless routing algorithms
for FiWi access networks are the so-called
delay-aware routing algorithm (DARA)~\cite{SYDM08},
delay-differentiated routing algorithm (DDRA)~\cite{CRSC10},
capacity and delay aware routing (CaDAR)~\cite{RRSG09},
and risk-and-delay aware routing (RADAR)
algorithm~\cite{SYDM07}. Recently proposed integrated routing
algorithms for path computation across the optical-wireless interface
include the so-called availability-aware routing~\cite{SYNC10}, multipath
routing~\cite{WWLQ10}, and energy-aware routing
algorithms~\cite{CTSM10}. Most of these previous studies formulated
routing in FiWi access networks as an optimization problem and
obtained results mainly by means of simulation.

In this paper, we present to the best of our knowledge the first
analytical framework that allows to evaluate the capacity and delay
performance of a wide range of FiWi network routing
algorithms and provides important
design guidelines for novel FiWi network routing algorithms that
leverage the different unique characteristics of disparate optical
fiber and wireless technologies. Although a few FiWi architectural
studies exist on the integration of EPON with
long-term evolution (LTE) (e.g.,~\cite{AEEH10})
or worldwide interoperability for microwave access (WiMAX)
wireless front-end networks
(e.g.,~\cite{ShTC07}), the vast
majority of studies, including but not limited to those mentioned in
the above paragraph, considered FiWi access networks consisting of a
conventional IEEE 802.3ah EPON fiber backhaul network and an IEEE
802.11b/g wireless local area network (WLAN)-based wireless mesh
front-end network~\cite{MaGh12}. Our framework encompasses not only
legacy EPON and WLAN networks, but also emerging next-generation
optical and wireless technologies, such as long-reach and multi-stage
10+ Gb/s time and/or wavelength division multiplexing (TDM/WDM) PONs
as well as Gigabit-class very high throughput (VHT) WLAN.

Our contributions are threefold. First, we develop a unified
analytical framework that comprehensively accounts for both optical
and wireless broadband access networking technologies. We note that
recent studies focused either on TDM/WDM
PONs only, 
e.g.,~\cite{AnFRR10,AuSRM10,AuSR11,BhSa10,LaVC07,Mar03,NgGB08,ZhAn11}, 
or on WLANs only, e.g.,~\cite{JaSi12}.
However, there is a need for a comprehensive
analytical framework that gives insights into the performance of
bimodal FiWi access networks built from disparate yet complementary
optical and wireless technologies. Toward this end, our framework is
flexibly designed such that it not only takes the capacity mismatch
and bit error rate differences between optical and wireless networks
into account, but also includes possible fiber cuts of optical (wired)
infrastructures.

Second, our analysis emphasizes
future and emerging next-generation PON and WLAN technologies, as
opposed to many previous studies that assumed
state-of-the-art solutions, e.g., conventional IEEE 802.11a WLAN
without frame aggregation~\cite{JaSi12}.
Our analytical approach in part builds on previous studies
and includes significant original analysis components to
achieve accurate throughput-delay modeling and cover the scope of
FiWi networks.
Specifically, we build on analytical models
of the distributed coordination function in
WLANs, e.g.,~\cite{Bianchi00,MaDL07}, and WLAN frame
aggregation, e.g.,~\cite{LiWo08}.
We develop an accurate delay model for multihop
wireless front-ends under nonsaturated and stable
conditions for traffic loads from both optical and wireless network
nodes, as detailed in Section~\ref{sec:wireless}.

Third, we verify our analysis by means of simulations and
present extensive numerical results to shed some light on the
interplay between different next-generation optical and wireless
access networking technologies and configurations for a variety of
traffic scenarios.
We propose an
\textit{optimized FiWi routing algorithm (OFRA)} based on our developed
analytical framework. The obtained results show that OFRA outperforms
previously proposed routing algorithms, such as
DARA~\cite{CRSC10}, CaDAR~\cite{RRSG09}, and RADAR~\cite{SYDM07}.
They also illustrate that it is key to
carefully select appropriate paths across the fiber backhaul in order
to minimize link traffic intensities and thus help stabilize the entire
FiWi access network.

To our best knowledge, the presented unified analytical
  framework is the first to allow capacity and
  delay evaluations of a wide range of FiWi network routing
  algorithms, both previously proposed and new ones.
  Our analytical framework covers not only legacy
  EPON and WLAN, but also next-generation high-speed long-reach WDM PON
  and emerging Gigabit-class VHT WLAN technologies.

The remainder of the paper is structured as follows. In
Section~\ref{sec:related}, we discuss related work and recent progress
on FiWi access networks. Section~\ref{sec:FiWi} describes FiWi access
networks based on next-generation PON and Gigabit-class WLAN
technologies in greater detail. Section~\ref{sec:model} outlines our
network model as well as traffic and routing assumptions. The capacity
and delay of the constituent fiber backhaul and wireless front-end
networks are analyzed in Sections~\ref{sec:fiber}
and~\ref{sec:wireless}, respectively, while the stability and
end-to-end delay of the entire FiWi access network are evaluated in
Section~\ref{sec:end-to-end}. Section~\ref{sec:results} presents
numerical and verifying simulation results.
Section~\ref{sec:conclusions} concludes the paper.

\section{Related Work}
\label{sec:related}
The recent survey of hybrid optical-wireless access networks~\cite{KWAA12}
explains the key
underlying photonic and wireless access technologies and describes
important FiWi access network architectures.
Energy-efficient FiWi network
architectures as well as energy-efficient medium access control (MAC)
and routing protocols were reviewed in~\cite{KaMo12}.
Recent efforts on energy-efficient routing in
FiWi access networks focused on routing
algorithms for cloud-integrated FiWi networks that
offload the wireless mesh front-end and
the optical-wireless gateways by placing cloud components, such as
storage and servers, closer to mobile end-users, while at the same time
maintaining low average packet delays~\cite{RRTM11,ReRT11}.
A delay-based admission control scheme for providing guaranteed
quality-of-service (QoS) in FiWi networks that deploy EPON as backhaul
for connecting multiple WiMAX base stations was studied in~\cite{DhHJ11}.

\begin{figure*}[t]
\begin{center}
\includegraphics[width=\textwidth]{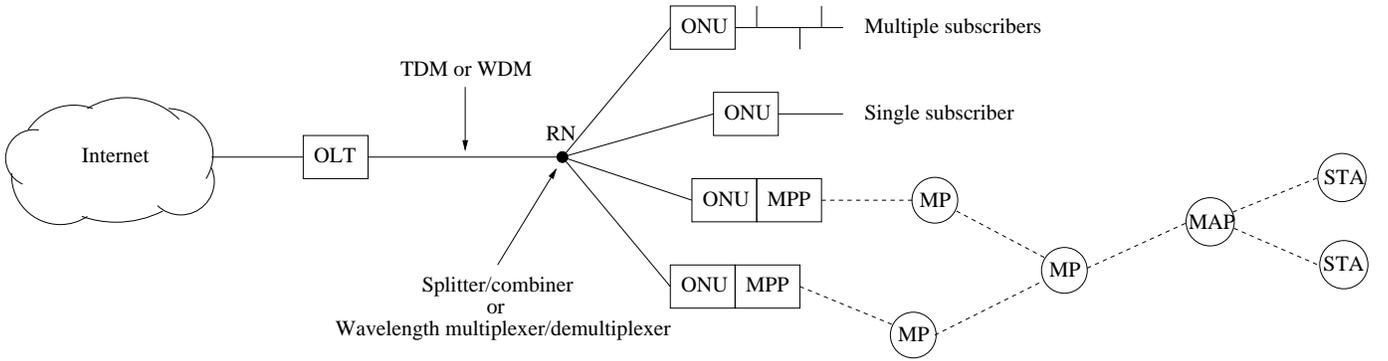}
\caption{FiWi network architecture based on single- or multi-stage TDM
  or WDM PON and multihop WMN.}
\label{fig:fig1}
\end{center}
\end{figure*}
A promising approach to increase throughput, decrease delay, and
achieve better load balancing and resilience is the use of multipath
routing schemes in the wireless mesh front-end of FiWi networks.
However, due to different delays along multiple paths, packets may
arrive at the destination out of order, which
deteriorates the performance of the Transmission Control Protocol
(TCP).
A centralized scheduling
algorithm at the optical line terminal (OLT) of an EPON that
resequences the in-transit packets of each flow to ensure in-order
packet arrivals at the corresponding destination
was examined in~\cite{LWQX11}. In addition,~\cite{LWQX11}
studied a dynamic bandwidth allocation (DBA) algorithm that
prioritizes flows that may trigger TCP's fast
retransmit and fast recovery, thereby further improving TCP
performance.

Given the increasing traffic amounts on FiWi networks, their
survivability has become increasingly important~\cite{Liu12,Maier12}.
Cost-effective protection schemes against link and node failures in the
optical part of FiWi networks have been proposed and optimized
in~\cite{FeRu11,KaMo10ICTON,Liu12aux,Yu09}.
The survivability of FiWi networks based on multi-stage PONs,
taking not only partial optical protection but also protection through
a wireless mesh network into account, was probabilistically analyzed
in~\cite{GhSM11}.
Deployment of both back-up fibers and radios was examined
in~\cite{Liu13}.

Recent research efforts have focused on the integration of
performance-enhancing network coding techniques to increase the
throughput and decrease the delay of FiWi access networks for unicast
and multicast traffic~\cite{FoMM11,ZhXW12}.

\section{FiWi Access Networks}
\label{sec:FiWi}
Most previous FiWi access
network studies considered a cascaded architecture consisting of a
single-stage PON and a multihop WMN, as shown in
Fig.~\ref{fig:fig1}. Typically, the PON is a
conventional IEEE 802.3ah compliant wavelength-broadcasting TDM EPON
based on a wavelength splitter/combiner at the remote node (RN), using
one time-shared wavelength channel for upstream (ONUs to OLT) transmissions  
and another time-shared wavelength channel for downstream (OLT to ONUs)
transmissions, both operating at a data rate of 1~Gb/s. 
A subset of ONUs may be located at the premises of residential
or business subscribers, whereby each ONU provides
fiber-to-the-home/business (FTTH/B) services to a single or multiple
attached wired subscribers.
Some ONUs have a mesh portal point (MPP) to interface with the WMN.
The WMN consists of mesh access
points (MAPs) that provide wireless FiWi network access to stations (STAs).
Mesh points (MPs) relay the traffic between MPPs and MAPs through
wireless transmissions.
Most previous FiWi studies assumed a WMN based on IEEE
802.11a/b/g WLAN technologies, offering a maximum raw data rate of 54
Mb/s at the physical layer.

Future FiWi access networks will leverage next-generation PON and
WLAN technologies to meet the ever increasing bandwidth requirements.
A variety of next-generation PON technologies are currently investigated
to enable short-term evolutionary and long-term revolutionary upgrades
of coexistent Gigabit-class TDM PONs~\cite{MaLI12}.
Promising solutions for PON evolution toward higher bandwidth per user
are ($i$) data rate upgrades to 10 Gb/s and higher, and ($ii$)
multi-wavelength channel migration toward wavelength-routing or
wavelength-broadcasting WDM PONs with or without cascaded
TDM PONs~\cite{AKWC11,KARC12}. Similarly, to alleviate the
bandwidth bottleneck of the wireless mesh front-end, future FiWi
networks are expected to be based on next-generation IEEE
802.11n WLANs, which offer data rates of 100 Mb/s or higher at the MAC
service access point, as well as emerging IEEE 802.11ac VHT WLAN
technologies that achieve raw data rates up to 6900 Mb/s.

\subsection{Next-Generation PONs}
\begin{figure}[t]
\begin{center}
\includegraphics[width=.5\textwidth]{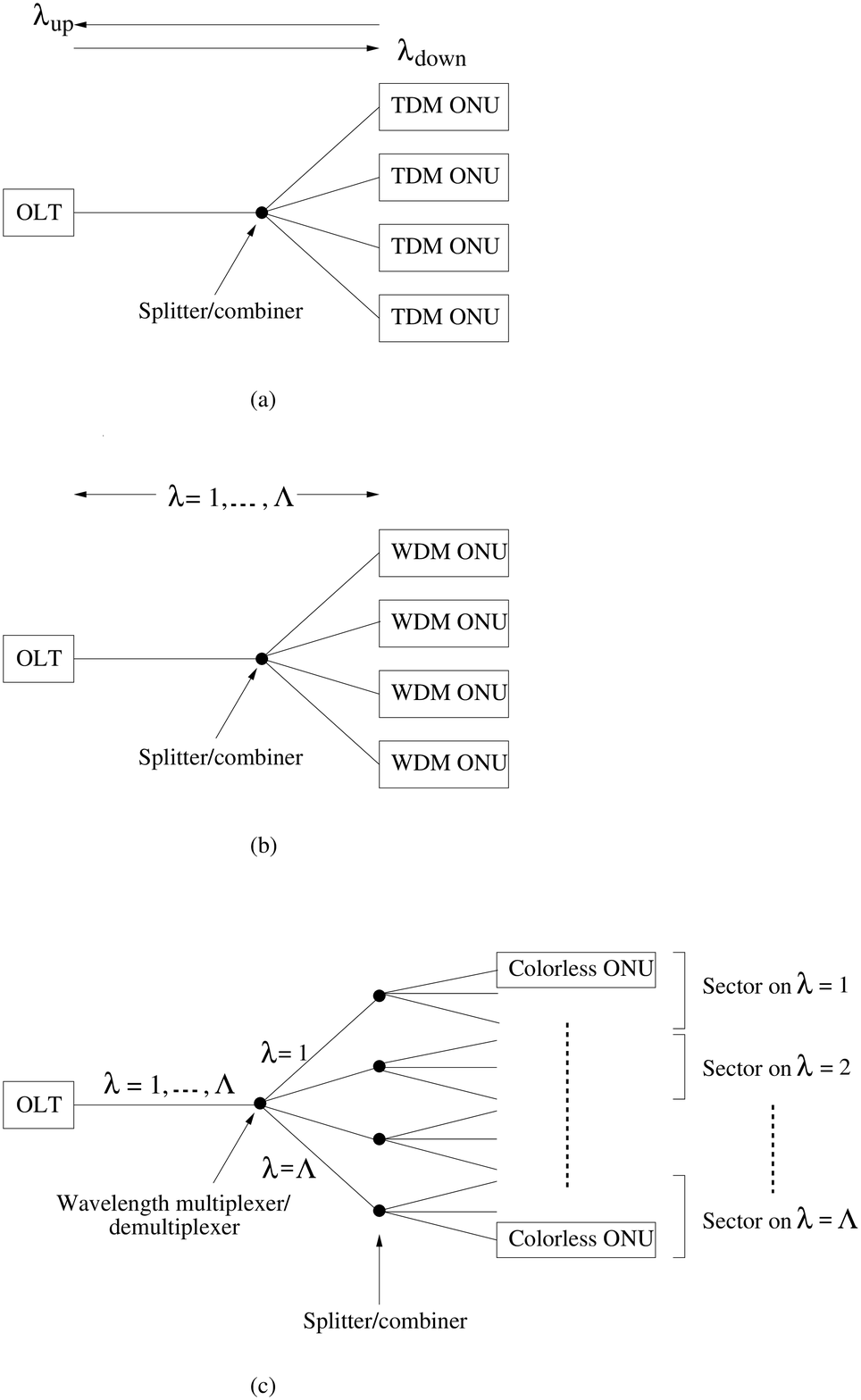}
\caption{Next-generation PONs: (a) High-speed TDM PON, (b)
  wavelength-broadcasting WDM PON, and (c) wavelength-routing
  multi-stage WDM PON.}
\label{fig:fig2}
\end{center}
\end{figure}
As shown in Fig.~\ref{fig:fig2}, current TDM PONs may evolve into
next-generation single- or multi-stage PONs of extended reach by
exploiting high-speed TDM and/or multichannel WDM technologies and
replacing the splitter/combiner at the RN with a wavelength
multiplexer/demultiplexer, giving rise to the following three types of
next-generation PONs:
\subsubsection{High-speed TDM PON}
Fig.~\ref{fig:fig2}(a) depicts a high-speed TDM PON, which maintains
the network architecture of conventional TDM PONs except that both the
time-shared upstream wavelength channel $\lambda_{\rm up}$ and
downstream wavelength channel $\lambda_{\rm down}$ and attached OLT
and TDM ONUs operate at data rates of 10 Gb/s or higher~\cite{Effe11}.
\subsubsection{Wavelength-broadcasting WDM PON}
A wavelength-broadcasting WDM PON has a splitter/combiner
at the RN and deploys multiple wavelength channels
$\lambda=1,\ldots,\Lambda$, as shown in Fig.~\ref{fig:fig2}(b). Each
of these $\Lambda$ wavelength channels is broadcast to all connected
WDM ONUs and is used for bidirectional transmission. Each WDM ONU
selects a wavelength with a tunable bandpass filter (e.g.,
fiber Bragg grating) and reuses the downstream modulated signal coming
from the OLT for upstream data transmission by means of remodulation
techniques, e.g., FSK for downstream and OOK for
upstream~\cite{GrEl08}.
\subsubsection{Wavelength-routing multi-stage WDM PON}
Fig.~\ref{fig:fig2}(c) shows a wavelength-routing WDM PON, where the
conventional splitter/combiner at the RN is replaced with a wavelength
multiplexer/demultiplexer, e.g., arrayed-waveguide grating (AWG), such
that each of the $\Lambda$ wavelength channels on the common feeder
fiber is routed to a different distribution fiber. A given wavelength
channel may be dedicated to a single ONU (e.g., business subscriber)
or be time shared by multiple ONUs (e.g., residential subscribers). In
the latter case, the distribution fibers contain one or more
additional stages, whereby each stage consists of a
wavelength-broadcasting splitter/combiner and each wavelength channel
serves a different sector, see Fig.~\ref{fig:fig2}(c).
Note that due to the wavelength-routing
characteristics of the wavelength multiplexer/demultiplexer, ONUs can
be made colorless (i.e., wavelength-independent) by using, for example,
low-cost reflective semiconductor optical amplifiers (RSOAs) that are
suitable for bidirectional transmission via
remodulation~\cite{AKWC11}. Wavelength-routing multi-stage
WDM PONs enable next-generation PONs with an extended optical range of
up to 100 km, thus giving rise to \textit{long-reach WDM PONs} at the
expense of additional in-line optical amplifiers. Long-reach WDM PONs
promise major cost savings by consolidating optical
access and metropolitan area networks~\cite{ShMi07}.

\subsection{Gigabit-Class WLAN}
IEEE 802.11n specifies a number of PHY and MAC enhancements for
next-generation WLANs. Applying orthogonal frequency division
multiplexing (OFDM) and multiple-input multiple-output (MIMO) antennas
in the PHY layer of IEEE 802.11n provides various capabilities, such as
antenna diversity (selection) and spatial multiplexing. Using multiple
antennas also provides multipath capability and increases both
throughput and transmission range. The enhanced PHY layer applies two
adaptive coding schemes: space time block coding (STBC) and low
density parity check coding (LDPC). IEEE 802.11n WLANs are able to
co-exist with IEEE 802.11 legacy WLANs, though in greenfield
deployments it is possible to increase the channel bandwidth from 20
MHz to 40 MHz via channel bonding, resulting in significantly
increased raw data rates of up to 600 Mb/s at the PHY layer.

\begin{figure}[t]
\begin{centering}
\setlength{\unitlength}{1mm}
\subfigure[\label{fig:3a}]{\includegraphics[width=0.475\textwidth]{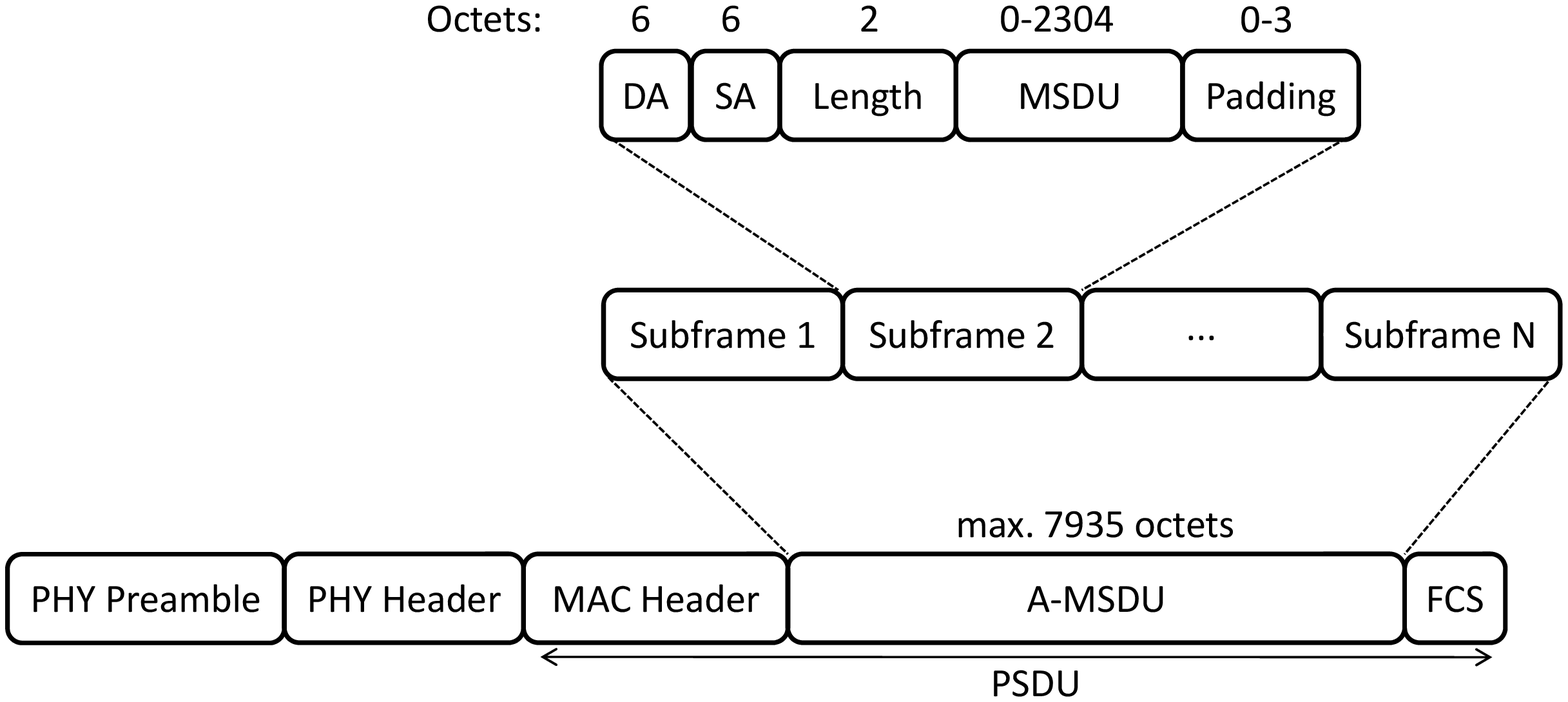}}
\subfigure[\label{fig:3b}]{\includegraphics[width=0.475\textwidth]{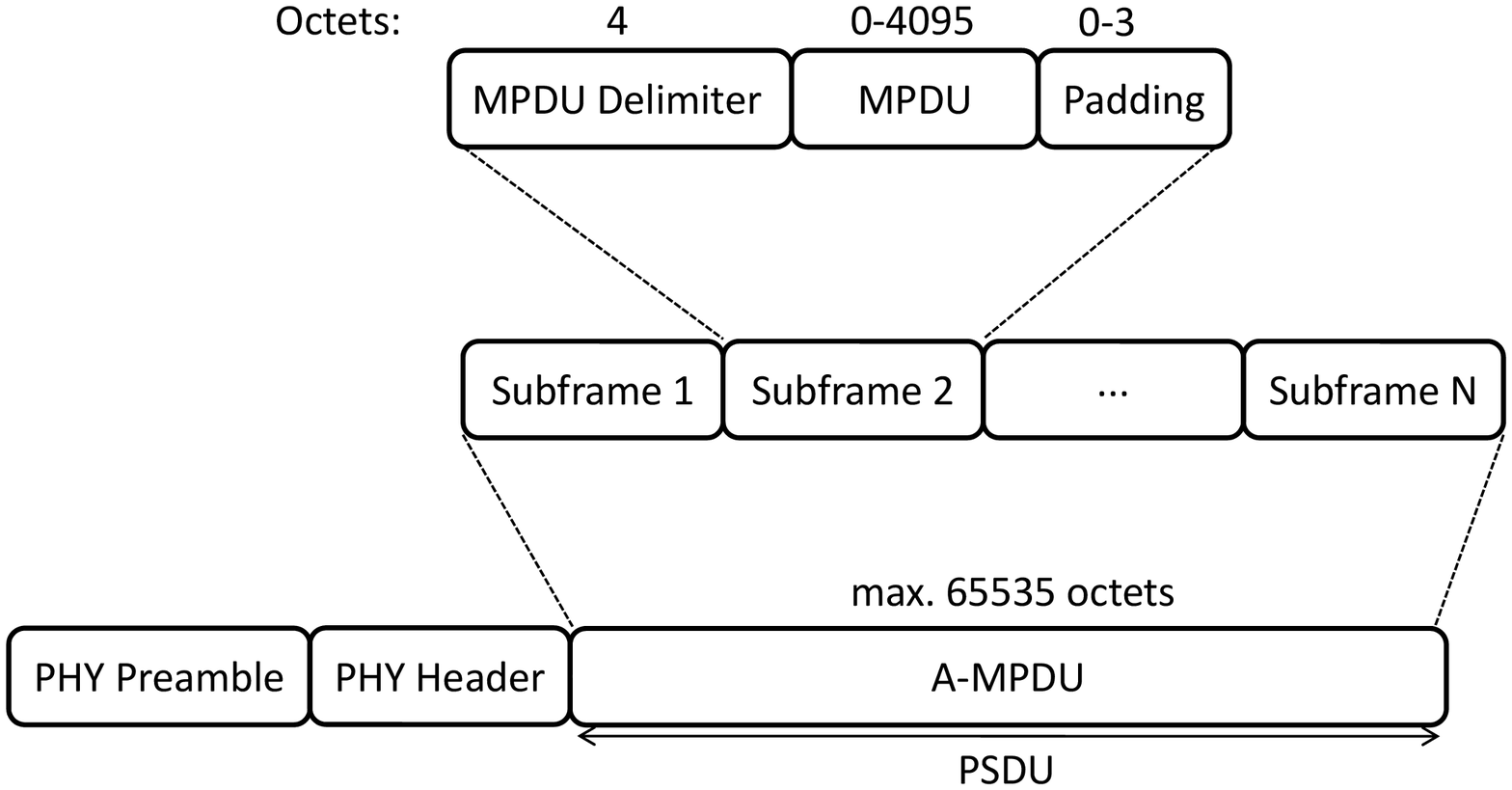}}
\caption{Frame aggregation schemes in next-generation WLANs: (a)
  A-MSDU, and (b) A-MPDU.}
\label{fig:fig3}
\end{centering}
\end{figure}
A main MAC enhancement of 802.11n is frame aggregation,
which comes in two flavors, as shown in Fig. \ref{fig:fig3}.

\noindent
{\bf Aggregate MAC Service Data Unit (A-MSDU):} Multiple MSDUs,
each up to 2304 octets long, are joined and encapsulated into a
separate subframe, see Fig.~\ref{fig:3a}. Specifically,
multiple MSDUs are packed into an A-MSDU, which is encapsulated into a
PHY service data unit (PSDU).  All constituent MSDUs must
have the same traffic identifier (TID) value (i.e., same QoS level)
and the resultant A-MSDU must not exceed the maximum size of 7935
octets. Each PSDU is prepended with a PHY preamble and PHY
header. Although the fragmentation of MSDUs with the same destination
address is allowed, A-MSDUs must not be fragmented.

\noindent
{\bf Aggregate MAC Protocol Data Unit (A-MPDU):}
Multiple MPDUs, each up to 4095 octets long, are joined
and inserted in a separate
subframe, see Fig.~\ref{fig:3b}. Specifically, multiple MPDUs are
aggregated into one PSDU of a maximum size 65535
octets. Aggregation of multiple MPDUs with different TID values into
one PSDU is allowed by using multi-TID block acknowledgment (MTBA).

Both A-MSDU and A-MPDU require only a single PHY preamble
and PHY header. In A-MSDU, the PSDU includes a single MAC header and
frame check sequence (FCS), as opposed to A-MPDU where each MPDU
contains its own MAC header and FCS. A-MPDU and A-MSDU can be used
separately or jointly.

Future Gigabit-class WMNs may be upgraded with emerging IEEE 802.11ac
VHT WLAN technologies that exploit further PHY enhancements to achieve
raw data rates up to 6900 Mb/s and provide an increased maximum
A-MSDU/A-MPDU size of 11406/1048575 octets~\cite{OKAC11}.

\section{Network Model}
\label{sec:model}
\subsection{Network Architecture}
\label{sec:netarch} We consider a PON consisting of one OLT and $O$
attached ONUs. The TDM PON carries one upstream wavelength channel
and a separate downstream wavelength channel. We suppose that both
the wavelength-broadcasting and the wavelength-routing multi-stage
WDM PONs carry $\Lambda$ bidirectional wavelength channels $\lambda
= 1, \ldots, \Lambda$. In the wavelength-routing multi-stage WDM
PON, the $O$ ONUs are divided into $\Lambda$ sectors. We use
$\lambda$ to index the wavelength channel as well as the
corresponding sector. In our model, sector $\lambda,\ \lambda
=1,\ldots, \Lambda$, accommodates $O_\lambda$ ONUs. Specifically,
ONUs with indices $o$ between $\sum_{\upsilon =
1}^{\lambda-1}O_\upsilon$ and
$\sum_{\upsilon=1}^{\lambda}O_\upsilon$ belong to sector $\lambda$,
i.e., form the set of nodes
\begin{equation}
\mathcal{S}_\lambda :=\left\{o|\sum_{\upsilon=1}^{\lambda-1}O_\upsilon < o \leq
       \sum_{\upsilon=1}^{\lambda}O_\upsilon \right\}.
\end{equation}
Thus, sector $\lambda = 1$ comprises
ONUs $o \in \mathcal{S}_1 = \{1,\ldots, O_1\}$,
sector $\lambda = 2$
comprises ONUs $o \in \mathcal{S}_2 = \{O_1+1,\ldots, O_1+O_2\}$,
and so on, while we assign the index $o=0$ to the OLT.
The one-way propagation delay between OLT
and ONUs of sector $\lambda$ is $\psi^{(\lambda)}$ (in seconds)
and the data rate of the associated wavelength channel $\lambda$ is
denoted by $c^{(\lambda)}$ (in bit/s).
Hence, each sector of
the wavelength-routing multi-stage WDM PON is allowed to operate at a
different data rate serving a subset of ONUs located at a different
distance from the OLT (e.g., business vs. residential service areas).
For ease of exposition, we assume that in the
wavelength-broadcasting TDM and WDM PONs all wavelength channels
operate at the same data rate $c$ (in bit/s)
and that all ONUs have the
one-way propagation delay $\psi$ (in seconds) from the OLT.

All or a subset of the $O$ ONUs are equipped with an MPP to interface
with the WMN.
The WMN is composed of different zones $z$, whereby each
zone operates on a distinct frequency such that the frequencies of
neighboring zones do not overlap. Frequencies may be spatially reused in nonadjacent zones.
A subset of MPs are assumed to be
equipped with multiple radios
to enable them to send and receive data
in more than one zone and thereby serve as relay nodes between
adjacent zones.
We denote each radio operating in a given relay MP in a given zone $z$
by a unique $\omega$.
The remaining MPs as well as all MPPs, MAPs, and STAs
are assumed to have only a single radio $\omega$ operating on the frequency of
their corresponding zone.
All wireless nodes are assumed to be stationary; incorporating
mobility is left for future research.
Adopting the notation proposed
in~\cite{DLLM06},
we let $\mathcal{R}_z$ denote the set of multi-radio
relay MPs and $\mathcal{L}_z$ denote the set of single-radio MPs,
MPPs, MAPs, and STAs in zone $z$.
Note that set $\mathcal{R}_z$ is empty if there are only
single-radio MPs in zone $z$.
Note that due to this set definition each
multi-radio MP is designated by multiple $\omega$;
one $\omega$ and corresponding set $R_z$ for each zone $z$
in which it can send and receive.
The WMN operates at a data rate $r$ (in bit/s).

In the WMN, we assume that the bit error rate (BER) of the wireless
channel is $p_b>0$. On the contrary, the BER of the PON is assumed to
be negligible and is therefore set to zero. However, individual fiber
links may fail due to fiber cuts and become unavailable for routing
traffic across the PON, as described next in more detail.
Throughout, we neglect nodal processing delays.

\subsection{Traffic Model and Routing}
\label{sec:traffic_model}
We denote $\mathcal{N}$ for the set of FiWi network nodes that act as
traffic sources and destinations. Specifically, we consider
$\mathcal{N}$ to contain the OLT,
the $O$ ONUs
(whereby a given ONU models the set of end users with wired access
to the ONU),
and a given number $N$ of STAs.
In our model, MPPs, MPs, and MAPs forward
in-transit traffic, without generating their own traffic. Hence, the
number of traffic sources/destinations is given by
$|\mathcal{N}| = 1 + O + N$.
Furthermore, we define the traffic
matrix $\mathbf{S} = (S_{ij}),\ i,j \in \mathcal{N}$, where $S_{ij}$ represents
the number of frames per second that are generated at FiWi network
node $i$ and destined to FiWi network node $j$ (note that $S_{ij}=0$
for $i=j$). We allow for any arbitrary distribution $F$ of the frame
length $L$ (in bit) and denote $\bar{L}$ and $\varsigma_{L}^{2}$ for the mean and
variance of the length of a frame, respectively. The
traffic generation is assumed to be ergodic and stationary.

Our capacity and delay analysis flexibly accommodates any
routing algorithm.
For each pair of FiWi network source node $i$ and
destination node $j$, a particular considered routing
algorithm results in a specific
traffic rate (in frames/s) $\Gamma_{ij}$ sent in the
fiber domain and traffic rate $\tilde{\Gamma}_{ij}$ sent in
the wireless domain. 
A conventional ONU $o$ without
an additional MPP cannot send in the wireless domain,
i.e., $\tilde{\Gamma}_{oj}=0$,
and sends its entire generated traffic to the OLT, i.e., $S_{oj}=\Gamma_{oj}$.
On the other hand, an ONU $o$ equipped
with an MPP can send in the wireless domain, i.e.,
$\tilde{\Gamma}_{oj}\geq 0$.
Note that we allow for multipath
routing in both the fiber and wireless domains, whereby traffic coming
from or going to the OLT may be sent across a single or multiple ONUs
and their collocated MPPs.
We consider throughout first-come-first-served
service in each network node.

\section{Fiber Backhaul Network}
\label{sec:fiber}
\subsection{Capacity Analysis}
For the wavelength-routing multi-stage WDM PON, we define the
normalized downstream traffic rate (intensity)
in sector $\lambda, \lambda = 1,\ldots, \Lambda$, as
\begin{equation}
\label{eq:WR_OLT}
\rho^{d,\lambda} := \frac{\bar{L}}{c^{(\lambda)}}
   \left( \sum_{o \in \mathcal{S}_\lambda} \Gamma_{0o} +
             \sum_{q = 1}^O \sum_{o \in \mathcal{S}_\lambda}\Gamma_{qo}\right),
\end{equation}
where the first term represents the traffic generated by the OLT
for sector $\lambda$ and the second term accounts for the traffic from all
ONUs sent to sector $\lambda$ via the OLT.
We define the upstream traffic rate (in frames/s) of ONU $o$ as
\begin{equation}
R_o^u:=\Gamma_{o0}+\sum_{q=1}^{O}\Gamma_{oq},
\end{equation}
where the first term denotes traffic destined to the OLT
and the second term represents the traffic sent to other ONUs via the OLT.
The normalized upstream traffic rate (intensity) of sector $\lambda$ is
\begin{equation}
\label{eq:WR_ONU}
\rho^{u,\lambda} := \frac{\bar{L}}{c^{(\lambda)}} \sum_{o \in \mathcal{S}_\lambda}R_o^u.
\end{equation}
For stability, the normalized downstream and upstream traffic rates have to satisfy
\begin{equation}
\label{eq:WR_1}
\rho^{d,\lambda} < 1 \mbox{ and } \rho^{u,\lambda} < 1
\end{equation}
in each sector $\lambda,\ \lambda = 1, \ldots, \Lambda$,
of the wavelength-routing multi-stage WDM PON.

In the wavelength-broadcasting TDM PON ($\Lambda=1$) and WDM PON
($\Lambda>1$), we define the upstream traffic intensity $\rho^u$ and downstream
traffic intensity $\rho^d$ as:
\begin{equation}
\label{eq:rho_u}
\rho^u:=\frac{\bar{L}}{\Lambda \cdot c}\sum_{o=1}^O\sum_{q=0}^O \Gamma_{oq}
\end{equation}
\begin{equation}
\label{eq:rho_d}
\rho^d:=\frac{\bar{L}}{\Lambda \cdot c}\sum_{q=0}^O\sum_{o=1}^O \Gamma_{qo}.
\end{equation}
The TDM and WDM PONs are stable if $\rho^u <1$ and $\rho^d <1$.
The delay analysis of Section~\ref{delay_ana:sec} applies only
for a stable network, which can be ensured through admission control
techniques.

\subsection{Delay Analysis}
\label{delay_ana:sec}
In the wavelength-routing multi-stage WDM PON,
the OLT sends a downstream frame to an ONU in sector $\lambda$ by
transmitting the frame on wavelength $\lambda$,
which is received by all ONUs in the sector.
We model all downstream transmissions in sector $\lambda$ to emanate from
a single queue.
For Poisson frame traffic, the downstream queueing delay is thus modeled
by an M/G/1 queue characterized by the
Pollaczek-Khintchine formula~\cite{Kleinrock75}
\begin{equation}
\Phi(\rho):=\frac{\rho}{2 c^{(\lambda)} (1-\rho)}
             \left(\frac{\varsigma_{L}^2}{\bar{L}}+\bar{L}\right)
\end{equation}
giving the total downstream frame delay
\begin{equation}
D^{d,\lambda}= \Phi\left(\rho^{d,\lambda}\right)
             +\frac{\bar{L}}{c^{(\lambda)}}+\psi^{(\lambda)}.
\end{equation}
Weighing the downstream delays $D^{d,\lambda}$ in the sectors $\lambda$
by the relative downstream traffic intensities
$\rho^{d,\lambda}/{\sum_{\lambda = 1}^\Lambda \rho^{d, \lambda}}$ in the sectors, gives the average downstream delay of the
wavelength-routing multi-stage WDM PON
\begin{equation}
\label{eq:WDM_PON_DSdelay}
D^d= \frac{1}{\sum_{\lambda = 1}^\Lambda \rho^{d, \lambda}} \sum_{\lambda=1}^\Lambda \rho^{d,\lambda}\cdot D^{d,\lambda}.
\end{equation}

For the upstream delay, we model each wavelength
channel $\lambda,\ \lambda = 1,\ldots, \Lambda$, as a single
upstream wavelength channel of a conventional EPON.
Accordingly, from Eq.~(39) in~\cite{ASHM08},
we obtain for the mean upstream delay of sector $\lambda$
\begin{equation}
D^{u,\lambda} = 2\psi^{(\lambda)} \cdot \frac{2-\rho^{u,\lambda}}{1-\rho^{u,\lambda}}
 +\Phi\left(\rho^{u,\lambda}\right)
       +\frac{\bar{L}}{c^{(\lambda)}}
\end{equation}
and the average upstream delay of the wavelength-routing
multi-stage WDM PON equals
\begin{equation}
D^u=\frac{1}{\sum_{\lambda = 1}^\Lambda \rho^{u, \lambda}}\sum_{\lambda=1}^\Lambda \rho^{u,\lambda}\cdot D^{u,\lambda}.
\end{equation}

To improve the accuracy of our delay analysis, we take into
account that traffic coming from an ONU $o$
in sector $\upsilon$ and destined to ONU $q$ in sector $\lambda$
is queued at the intermediate OLT before being sent
downstream to ONU $q$, i.e., the OLT acts like an insertion buffer
between ONUs $o$ and $q$. Consequently, to compensate for the queueing
delay at the OLT we apply the method proposed in~\cite{BuSc83} by
subtracting the correction term
\begin{equation}
B^{d,\lambda}=\sum_{\upsilon=1}^\Lambda \Phi\left(\rho^{\upsilon \to \lambda}\right),
\end{equation}
whereby for the setting that $c^{(\lambda)}=c$ for all channels $\lambda$
\begin{equation} \label{eq:rhost}
\rho^{\upsilon \to \lambda}=\frac{\bar{L}}{c}\cdot\sum_{o \in \mathcal{S}_\upsilon}
               \sum_{q \in \mathcal{S}_\lambda} \Gamma_{oq}
\end{equation}
denotes the rate of upstream traffic in sector $\upsilon$ destined for
sector $\lambda$, from the above calculated mean downstream delay.
Thus, for sector $\lambda,\ \lambda=1,\ldots, \Lambda$,
the corrected mean downstream delay $\tilde{D}^{d,\lambda}$ is given by
\begin{equation}
\tilde{D}^{d,\lambda}=D^{d,\lambda}-B^{d,\lambda}.
\end{equation}
By replacing $D^{d,\lambda}$ with $\tilde{D}^{d,\lambda}$ in
Eq. (\ref{eq:WDM_PON_DSdelay}) we obtain a more accurate calculation
of the average downstream delay for the wavelength-routing multi-stage
WDM PON, as examined in Section~\ref{sec:results}.

Next, we evaluate the average downstream and upstream delays for
the wavelength-broadcasting TDM PON ($\Lambda=1$) and WDM PON ($\Lambda>1$).
With the aforementioned correction term the average
downstream and upstream delays are given by
\begin{equation}
D^d=\Phi\left(\rho^d\right) + \frac{\bar{L}}{c} + \psi - B^d
\end{equation}
and
\begin{equation}
D^u=\Phi\left(\rho^u\right)+\frac{\bar{L}}{c} + 2 \psi \frac{2 - \rho^u}{1 - \rho^u} - B^u,
\end{equation}
respectively, whereby
\begin{equation}
B^d=B^u=\Phi\left(\frac{\bar{L}}{\Lambda\cdot c}
             \sum_{o=1}^O\sum_{q=1}^O \Gamma_{oq}\right).
\end{equation}

\section{Wireless Front-End Network}
\label{sec:wireless}

So far, we have analyzed only the optical fiber backhaul of the FiWi
network. Next, we focus on the wireless front-end.
In the following,
we derive multiple relations between known parameter values and
unknown variables. Afterwards, we outline how to obtain the
unknowns numerically.
More specifically, in
Sections~\ref{ftm:sec}--\ref{sec:durframetra} we build on and adapt
existing models of distributed coordination~\cite{Bianchi00,MaDL07,DLLM06}
and frame aggregation~\cite{LiWo08} in WLANs to formulate
the basic frame aggregate transmission and collision
probabilities as well as time slot duration in the distributed access system.
We note that these existing models have primarily focused
on accurately representing the collision probabilities
and system throughput; we found that directly adapting
these existing models gives delay characterizations that are
reasonably accurate only for specific scenarios, such as single-hop
networking, but are very coarse for multi-hop networking.
In Sections~\ref{sertime_fa:sec}--\ref{dWMNp:sec}
we develop a general multihop delay model that is simple, yet accurate
by considering the complete service time
of a frame aggregate in the wireless front-end network carrying
traffic streams from and to both wireless and optical network nodes.

\subsection{Frame Traffic Modeling}
\label{ftm:sec}
As defined in Section~\ref{sec:netarch}, we denote the radio
operating in a given STA or ONU equipped
with an MPP by a unique $\omega$.
Moreover, we denote each radio operating in a given relay MP in a
unique zone $z$ by a unique $\omega$.
For ease of exposition, we refer to ``radio $\omega$''
henceforth as ``node $\omega$.''

Similar to~\cite{DLLM06}, we model time
as being slotted and
denote $E_\omega$ for the mean duration of a time slot at node $\omega$.
The mean time slot duration $E_\omega$ corresponds to the average
time period required for a successful frame transmission,
a collided frame transmission, or an idle waiting slot
at node $\omega$ and is evaluated in Section~\ref{sec:durframetra}.
We let $q_\omega$ denote the
probability that there is a frame waiting for transmission at node
$\omega$ in a time slot.

For an STA or ONU with collocated MPP $\omega$ we
denote $\sigma_\omega$ for the traffic load that emanates from
node $\omega$, i.e.,
\begin{equation}
\sigma_\omega:=\sum_{\forall{i}} \tilde{\Gamma}_{\omega i}.
\end{equation}
For a relay MP we obtain
for a given wireless mesh routing algorithm the frame arrival rate
for each of the MP's radios $\omega \in \mathcal{R}_z$
associated with a different zone $z$:
\begin{equation}
\sigma_\omega:=\sum_{\forall{i,j}} \tilde{\Gamma}_{ij},
\end{equation}
whereby $i$ and $j$ denote any pair of STA or ONU with collocated MPP
that send traffic on a path via relay MP $\omega$, as computed by the given
routing algorithm for the wireless mesh front-end of the FiWi
network.

For exponentially distributed inter-frame arrival
times with mean $1/\sigma_\omega$ (which occur for a Poisson process with
rate $\sigma_\omega$), $q_\omega$ is related to the offered frame
load at node $\omega$ during mean time slot duration $E_\omega$ via
\begin{equation}
\label{eq:q_omega}
1-q_\omega=e^{-\sigma_\omega\cdot E_\omega}.
\end{equation}

\subsection{Frame Aggregate Error Probability}
In this section, we first characterize the sizes of the frame aggregates
and then the frame aggregate error probability.
For a prescribed distribution $F(l)$ of the size (in bit) of a single frame,
e.g., the typical trimodal IP packet size distribution,
the distribution $A(l)$ of the size (in bit)
of a transmitted A-MSDU or A-MPDU
can be obtained as the convolution of $F$ with itself,
i.e.,
\begin{equation}
A(l)=(F*F*\ldots *F)(l).
\end{equation}
The number of required convolutions equals the number of frames
carried in the aggregate, which in turn depends on the minimum frame
size, including the MAC-layer overhead of the corresponding frame
aggregation scheme,
and the maximum size of an A-MSDU/A-MPDU $A_{\max}^{\text{A-MSDU/A-MPDU}}$ (see Fig.~\ref{fig:fig3}).
From the distribution $A(l)$ we obtain
the average frame aggregate sizes $E[\text{A-MSDU}]$ and $E[\text{A-MPDU}]$.
Correspondingly, we divide the traffic rate $\tilde{\Gamma}_{i j}$
(in frames/s) by the average number of frames in an aggregate
to obtain the traffic rate in frame aggregates per second.

Moreover, as ground work for Section~\ref{sec:durframetra}
we obtain
the average size of the longest A-MSDU,  $E[\text{A-MSDU}^*]$,
and longest A-MPDU, $E[\text{A-MPDU}^*]$,  involved in a collision
with the simplifying assumption of neglecting the
collision probability
of more than two packets~\cite{Bianchi00} as
\begin{equation}
\label{eq:max_aggregate}
E[\text{A-MSDU}^*/\text{A-MPDU}^*] =
          \int_0^{A_{\max}^{\text{A-MSDU/A-MPDU}}}\left(1-A(x)^2\right) dx.
\end{equation}
The probability $p_e$
of an erroneously transmitted frame aggregate,
referred to henceforth as ``transmission error'',
can be evaluated in terms of bit error probability $p_b$
and size $A$ of a transmitted A-MSDU
(with distribution $A(l)$)
with~\cite[Eqn.~(16)]{Lin06};
for A-MPDU, $p_e$ can be evaluated
in terms of $p_b$ and the sizes $L_i$ of the aggregated frames
with~\cite[Eqn.~(18)]{Lin06}.
In particular,
\begin{equation}
p_e=\left\{
\begin{array}{ll}
1-(1-p_b)^A & \text{for A-MSDU},\\
\prod_i\left(1-(1-p_b)^{L_i}\right) & \text{for A-MPDU},
\end{array}\right.
\end{equation}
where $A$ is the size $A$ of a transmitted A-MSDU
(with distribution $A(l)$), index $i$ runs from one
to the total number of aggregated frames, and $L_i$ is the size of the
$i$th frame in a transmitted A-MPDU.

\subsection{Probabilities for Frame Aggregate Collision and Successful Frame Aggregate Transmission}
Following~\cite{DLLM06}, we note that the transmission of
any transmitting node $\omega \in \mathcal{R}_z \cup \mathcal{L}_z$
in zone $z$ cannot collide if none of the other nodes
$\nu \in \mathcal{R}_z \cup \mathcal{L}_z, \nu \neq \omega$
transmits, i.e., we obtain the collision probability $p_{c, \omega}$ as
\begin{equation}
\label{eq:p_c}
1-p_{c, \omega} =
   \prod_{\substack{\nu \in \mathcal{R}_z\cup\mathcal{L}_z \\ \nu \neq \omega}} (1-\tau_\nu),
\end{equation}
where $\tau_\nu$ denotes the transmission probability of WMN node $\nu$.
Note that if the considered node is a relay MP,
Eq.~(\ref{eq:p_c}) holds for
each associated zone $z$ (and corresponding radio $\omega$).
We define the probability of either a collision or transmission
error $p_\omega$,
in brief collision/transmission error probability, as
\begin{equation}
\label{eq:p_omega}
1-p_\omega = (1-p_e)\cdot (1 - p_{c, \omega}).
\end{equation}
The transmission probability $\tau_\omega$ for
any node $\omega \in\mathcal{R}_z \cup \mathcal{L}_z$
can be evaluated as a function of
the frame waiting probability $q_{\omega}$,
the frame collision/transmission error probability $p_{\omega}$,
the minimum contention window $W_0$, and the maximum
backoff stage $H$ by~\cite[Eqn.~(1)]{DLLM06},
as explained in~\cite{MaDL07}.
In particular,
\begin{equation}
\label{eq:tau_omega}
\tau_\omega=\frac{1}{\eta}\left(\frac{q_\omega^2\cdot W_0}{(1-q_\omega)
(1-p_\omega)[1-(1-q_\omega)^{W_0}]}-\frac{q_\omega^2(1-p_\omega)}
{1-q_\omega}\right),
\end{equation}
with
\begin{eqnarray}
\eta \!\!\! &=& \!\!\! \frac{q_\omega W_0}{1-(1-q_\omega)^{W_0}}
    +\frac{q_\omega W_0 (q_\omega W_0+3q_\omega-2)}{2(1-q_\omega)
             [1-(1-q_\omega)^{W_0}]}\nonumber \\
&& \ \ \ + (1-q_\omega) \nonumber \\
&& \  \ \ +\frac{q_\omega(W_0+1)[p_\omega(1-q_\omega)-q_\omega(1-p_\omega)^2]}
       {2(1-q_\omega)}\nonumber \\
&& \ \ \ +\frac{p_\omega q_\omega^2}{2(1-q_\omega)(1-p_\omega)}
        \left(\frac{W_0}{1-(1-q_\omega)^{W_0}}-(1-p_\omega)^2\right)
\nonumber \\
&& \ \ \ \ \ \ \ \ \ \ \
   \ \cdot \left( \frac{2W_0 [1-p_\omega-p_\omega(2p_\omega)^{H-1}]}
{1 - 2p_\omega}  +1 \right),
\end{eqnarray}
where $W_0$ is node $\omega$'s minimum contention window,
$W_0 2^H$ is the node's maximum window size,
and $H$ is the maximum backoff stage.

The probability that there is at least one
transmission taking place in zone $z$ in a given time slot
is given by
\begin{equation}
P_{tr,z}=1-\prod_{\omega \in \mathcal{R}_z \cup \mathcal{L}_z} (1-\tau_\omega).
\end{equation}
A successful frame aggregate transmission occurs
if exactly one node $\omega$ transmits
(and all other nodes $\nu \neq \omega$ are silent),
given that there is a transmission, i.e.,
\begin{equation}
P_{s,z}= \frac{1}{P_{tr,z}}
    \left(\sum_{\omega \in \mathcal{R}_z \cup \mathcal{L}_z}\tau_\omega\cdot
       \prod_{\substack{\nu \in \mathcal{R}_z \cup \mathcal{L}_z
                 \\ \nu \neq \omega}}(1-\tau_\nu)\right).
\end{equation}

\subsection{Duration of Single Frame Aggregate Transmission}
\label{sec:durframetra}
We denote $\epsilon$ for the duration of an empty time slot without
any data transmission on the wireless channel in zone $z$,
which occurs with probability $1 - P_{tr,z}$.
With probability $P_{tr,z}$ there is a transmission in a given
time slot in zone $z$, which is
successful with probability $P_{s,z}$ and unsuccessful
(resulting in a collision) with
the complementary probability $1 - P_{s,z}$.

We denote $T_{s,z}$ for the mean duration of a successful
frame aggregate transmission and $T_{c,z}$ is the mean duration
of a frame aggregate transmission with collision in zone $z$.
Note that $T_{s,z}$ and $T_{c,z}$ depend
on the frame aggregation technique (A-MSDU or A-MPDU) and
 on the access mechanism $\alpha$ (basic access denoted
by $\alpha = {\rm basic}$ or RTS/CTS denoted by $\alpha = {\rm RTS/CTS}$).
For the basic access mechanism,
we define
$\Theta_s^{\rm basic} = \text{DIFS} + \text{PHY Header} + \text{SIFS}
+ \delta + \text{ACK}/r  + \delta$,
where $\delta$ denotes the propagation delay
and $r$ the WMN data rate.
For the RTS/CTS access mechanism,
we define
$\Theta_s^{\rm RTS/CTS} = 
\text{DIFS} + \text{RTS}/r + \text{SIFS}
+ \delta + \text{CTS}/r +
\text{SIFS} + \delta + \text{PHY Header} +
\text{SIFS} + \delta + \text{ACK}/r +  \delta$.
(Note that in IEEE 802.11n the parameters ACK, RTS, and CTS
as well as the PHY/MAC Header and FCS below are given in bits,
while the other parameters are given in seconds.) Then, for a successful frame aggregate transmission we have:
\begin{equation}
T_{s,z}^{\alpha}= \left\{
\begin{array}{ll}
\Theta_s^{\alpha} + (\text{MAC Header}\ +\\ E[\text{A-MSDU}] + \text{FCS})/r
      & \text{for A-MSDU}\\
\\
\Theta_s^{\alpha} + E[\text{A-MPDU}]/r\ & \text{for A-MPDU}.
\end{array}\right.
\end{equation}
Moreover, with
$\Theta_c^{\rm basic} = \text{PHY Header} + \text{DIFS} + \delta$,
for a collided frame aggregate transmission we have:
\begin{equation}
T_{c,z}^{\rm basic}= \left\{
\begin{array}{ll}
\Theta_c^{\rm basic} + (\text{MAC Header}\ +\\ E[\text{A-MSDU}^*] + \text{FCS})/r
& \text{for A-MSDU},\\
\\
\Theta_c^{\rm basic} + E[\text{A-MPDU}^*]/r\  & \text{for A-MPDU}
\end{array}\right.
\end{equation}
as well as for both A-MSDU and A-MPDU,
\begin{equation}
T_{c,z}^{\rm RTS/CTS} = \text{RTS}/r + \text{DIFS} + \delta.
\end{equation}

Thus, we obtain the expected time slot duration $E_\omega$
at node $\omega$ in zone $z$ of our network model
(corresponding to \cite[Eq.~(13)]{Bianchi00}) as
\begin{eqnarray}
\label{eq:E_omega}
E_\omega=(1-P_{tr,z})\epsilon + P_{tr,z}
          \left[ P_{s,z}T_{s,z}^\alpha +  (1 - P_{s,z}) T_{c,z}^\alpha \right].
\end{eqnarray}

Equations (\ref{eq:q_omega}), (\ref{eq:p_omega}),~\cite[Eqn.~(1)]{DLLM06},
and (\ref{eq:E_omega})
can be solved numerically for the unknown variables
$q_\omega$, $p_\omega$, $\tau_\omega$, and $E_\omega$
for each given set of values for the known network model parameters.
We use the obtained numerical solutions to evaluate the mean delay
at node $\omega$ as analyzed in the following
Sections~\ref{sertime_fa:sec} and~\ref{wmnnodedel:sec}.

\subsection{Service Time for Frame Aggregate}
\label{sertime_fa:sec}
We proceed to evaluate the expected service (transmission) time for a
frame aggregate, which may require several transmission attempts,
at a given node $\omega$.
With the basic access mechanism, the transmission of the
frame aggregate occurs without a collision ($j = 0$) or transmission error
with probability $1 - p_\omega$ (\ref{eq:p_omega}), requiring
one $T_{s,z}^{\mathrm{basic}}$.
With probability $p_\omega^j (1 - p_\omega)$,
the frame aggregate suffers $j,\ j = 1, 2, \ldots $, collisions
or transmission errors, requiring $j$ backoff procedures
and re-transmissions.
Thus, the expected service time for basic access is
\begin{eqnarray}
\Delta_{\mathrm{ser}, \omega}^{\mathrm{basic}}  =
           \ \ \ \ \ \ \ \ \ \ \ \ \ \ \ \ \ \ \ \ \ \ \ \ \ \ \ \ \ \ \ \
 \ \ \ \ \ \ \ \ \ \ \ \ \ \ \ \ \ \ \ \ \ \ \ \ \ \ \ \ \ \ \nonumber\\
     \sum_{j=0}^{\infty} p_\omega^j (1 - p_\omega )
          \left( j T_{c, z}^{\mathrm{basic}}+
             \sum_{b=1}^{j} \frac{2^{\min(b, H)}W_0 - 1}{2} \epsilon \right )
               + T_{s, z}^{\mathrm{basic}}.
\end{eqnarray}

For the RTS/CTS access mechanism, collisions can occur only for the
RTS or CTS frames (which are short and have negligible probability
of transmission errors), whereas transmission errors may occur for
the frame aggregates. Collisions require only retransmissions of the
RTS frame, whereas transmission errors require retransmissions of
the entire frame aggregate. More specifically, only one frame
transmission ($k = 1$) is required if no transmission error occurs;
this event has probability $1 - p_e$. This transmission without
transmission error may involve $j,\ j = 0, 1, 2, \ldots$, collisions
of the RTS/CTS frames. On the other hand, two frame transmissions
($k = 2$) are required if there is once a transmission error; this
event has probability $p_e(1-p_e)$. This $k = 2$ scenario requires
twice an RTS/CTS reservation, which each time may experience $j, j =
0, 1, 2, \ldots$ collisions, as well as two full frame transmission
delays $T_{s,z}$. Generally, $k,\ k = 1, 2, \ldots$, frame
transmissions are required if $k-1$ times there is a frame
transmission error. Each of the $k$ frame transmission attempts
requires an RTS/CTS reservation and a full frame transmission delay
$T_{s,z}$. In summary, we evaluate the mean service delay for a
frame aggregate with RTS/CTS access as
\begin{eqnarray}
\Delta_{\mathrm{ser}, \omega}^{\mathrm{RTS/CTS}} \!\!\!& = &\!\!\!
\sum_{k=1}^{\infty} p_e^{k-1} (1 - p_{e}) k \left[
   \sum_{j=0}^{\infty} p_{c,\omega}^j (1 - p_{c, \omega})
  \right . \nonumber \\
     &&  \!\!\!\!\!\!\!\!\!\!\!\!\!\!\!\!\!\!\!\!\!\!\!\!\!\!\!\!\!
\!\!\!\!\!\!\!\!\!\!
\left .
    \left( \sum_{b=1}^{j} \frac{2^{\min(b, H)}W_0 - 1}{2} \epsilon +
            j T_{c, z}^{\mathrm{RTS/CTS}} \right)
  + T_{s,z}^{\mathrm{RTS/CTS}}  \right].
\end{eqnarray}

\subsection{Delay at WMN Node}
\label{wmnnodedel:sec}
We first evaluate the overall service time $\Delta_\omega$
from the time instant when a
frame aggregate arrives at the head of the queue
at node $\omega$ to the completion of its successful transmission.
Subsequently, with $\Delta_\omega$ characterizing the
overall service time at node $\omega$, we evaluate the
queueing delay $D_\omega^{\rm wi}$.

The overall service time $\Delta_\omega$ is given by the service time
$\Delta_{\mathrm{ser}, \omega}^\alpha$ required for transmitting a frame aggregate
and the sensing delay $\Delta_{\mathrm{sen}, \omega}$ required for
the reception of frame aggregates by node $\omega$ from other
nodes, i.e.,
\begin{equation} \label{Delom_ovall:eqn}
\Delta_{\omega} =  \Delta_{\mathrm{ser}, \omega}^{\alpha}
          + \Delta_{\mathrm{sen}, \omega}.
\end{equation}
As a first step towards modeling the sensing delay at a node $v$, we
consider the service times $\Delta_{\mathrm{ser}, v_1}^\alpha$ at nodes
$v_1 \neq \nu$ and scale these service times linearly with the
corresponding traffic intensities 
$\sigma_{v_1}/(1 / \Delta_{\mathrm{ser}, v_1}^\alpha)$
to obtain the sensing delay component
\begin{equation}
D_{\mathrm{sen}, \nu} = \sum_{\forall_{v_1 \neq \nu \ \mathrm{ in } z}}
  \frac{\sigma_{v_1}}{1/\Delta_{\mathrm{ser}, v_1}^{\alpha}} 
       \Delta_{\mathrm{ser}, v_1}^{\alpha}.
\end{equation}
As a second modeling step, we consider the service times plus
sensing delay components scaled by the respective
traffic intensities to obtain the sensing delay
\begin{equation}
\Delta_{\mathrm{sen}, \omega} = \sum_{\forall_{\nu \neq \omega \mathrm{ in } z}}
    \frac{\sigma_{\nu}}{1/(\Delta_{\mathrm{ser}, \nu}^{\alpha}
        + D_{\mathrm{sen}, \nu})} (\Delta_{\mathrm{ser}, \nu}^{\alpha}
          + D_{\mathrm{sen}, \nu})
\end{equation}
employed in the evaluation of the overall service
delay (\ref{Delom_ovall:eqn}).


We approximate the queue at node $\omega$ by an
M/M/1 queue with mean arrival rate $\sigma_\omega$ and mean
service time $\Delta_\omega$. This queue is stable if
\begin{equation}
\label{eq:rho_omega}
  \sigma_\omega \cdot \Delta_\omega  < 1.
\end{equation}
The total delay (for queueing plus service)
at node $\omega$ is then given by
\begin{equation}
D_\omega^{\rm wi} = \frac{1}{\frac{1}{\Delta_\omega} - \sigma_\omega}.
\end{equation}

If node $\omega$ is an ONU with a collocated MPP
the accuracy of the queueing delay calculation is improved
by subtracting a correction term:
\begin{equation}
\tilde{D}_\omega^{\rm wi} = D_\omega^{\rm wi}
  - \Phi\left( \frac{\bar{L}}{c} \sum_{\forall i,j} S_{ij} \right)
\end{equation}
for the wavelength-broadcasting TDM PON and WDM PON, or
\begin{equation}
\tilde{D}_\omega^{\rm wi} = D_\omega^{\rm wi}
  - \Phi\left( \frac{\bar{L}}{c^{(\lambda)}} \sum_{\forall i,j} S_{ij} \right)
\end{equation}
for the wavelength-routing multi-stage WDM PON, whose sector $\lambda$ accommodates the ONU with collocated MPP. Note that $\frac{\bar{L}}{c}\sum_{\forall i,j} S_{ij}$ or $\frac{\bar{L}}{c^{(\lambda)}} \sum_{\forall i,j} S_{ij}$ accounts for the traffic of all pairs of source node $i$ and destination node $j$ traversing ONU $\omega$ from the fiber backhaul towards the wireless front-end network.

\subsection{Delay on WMN Path}
\label{dWMNp:sec}
In order to obtain the delay in the wireless front-end of our FiWi
network, we have to average the
sums of the nodal delays of all possible paths for all
pairs of source node $i$ and destination node $j$:
\begin{equation}
D^{\rm wi} = \sum_{i,j}\frac{\tilde{\Gamma}_{ij}}{\sum_{i,j}\tilde{\Gamma}_{ij}}
        \left(\sum_{\substack{\forall \omega \text{ on path}\\\text{from $i$ to $j$}}}
             \left(D_\omega^{\rm wi}
                          - B_{ij\omega}^{\rm wi} \right)\right),
\end{equation}
with the queueing delay correction terms
\begin{equation}
B_{ij\omega}^{\rm wi}=\frac{\tilde{\Gamma}_{ij} \cdot \Delta_\omega}
     { \frac{1}{\Delta_\omega} - \tilde{\Gamma}_{ij}},
\end{equation}
whereby $\tilde{\Gamma}_{ij} \cdot \Delta_\omega$
is the traffic intensity at node $\omega$ due to traffic
flowing from source node $i$ to destination node $j$.

\section{FiWi Network Stability and Delay}
\label{sec:end-to-end}
The entire FiWi access network is stable if and
only if all of its optical and wireless subnetworks are stable. If the
optical backhaul consists of a wavelength-routing multi-stage WDM PON
the stability conditions in Eq.~(\ref{eq:WR_1})
must be satisfied. In the case of the wavelength-broadcasting TDM and
WDM PON, the optical backhaul is stable if both $\rho^u$ and $\rho^d$
defined in Eqs.~(\ref{eq:rho_u}) and~(\ref{eq:rho_d}),
respectively, are smaller than one.
The wireless mesh front-end is stable if the stability condition
in Eq.~(\ref{eq:rho_omega})
is satisfied for each WMN node.

We obtain the mean end-to-end delay of the entire bimodal FiWi
access network as
\begin{equation}
\label{eq:delay}
D=D^d + D^u + D^{\rm wi}.
\end{equation}

\section{Numerical and Simulation Results}
\label{sec:results}
We set the parameters of the FiWi mesh
front-end to the default values for next-generation WLANs~\cite{PeSt08},
see Table~\ref{tab:1}.
We consider a distance of 1 km
between any pair of adjacent WMN nodes
(which is well within the maximum
distance of presently available outdoor wireless access points),
translating into a propagation delay of $\delta=1/3\cdot 10^{-5}$ s.
\begin{table}[t]
\caption{FiWi Network Parameters}
\label{tab:1}
\begin{center}
\begin{tabular}{|c|c|}
\hline
Parameter & Value \\
\hline\hline
Min.\ contention window $W_0$ & 16 \\
\hline
Max. backoff state $H$ & 6 \\
\hline
Empty slot duration $\epsilon$ & 9 $\mu$s\\
\hline
SIFS & 16 $\mu$s\\
\hline
DIFS & 34 $\mu$s\\
\hline
PHY Header & 20 $\mu$s\\
\hline
MAC Header & 36 bytes\\
\hline
RTS & 20 bytes\\
\hline
CTS & 14 bytes\\
\hline
ACK & 14 bytes\\
\hline
FCS & 4 bytes\\
\hline
\end{tabular}
\end{center}
\label{default}
\end{table}

\subsection{Model Verification}
We first verify the accuracy of our probabilistic analysis by means of 
simulations.
\subsubsection{Configuration}
\begin{figure}[t]
\begin{center}
\includegraphics[width=.5\textwidth]{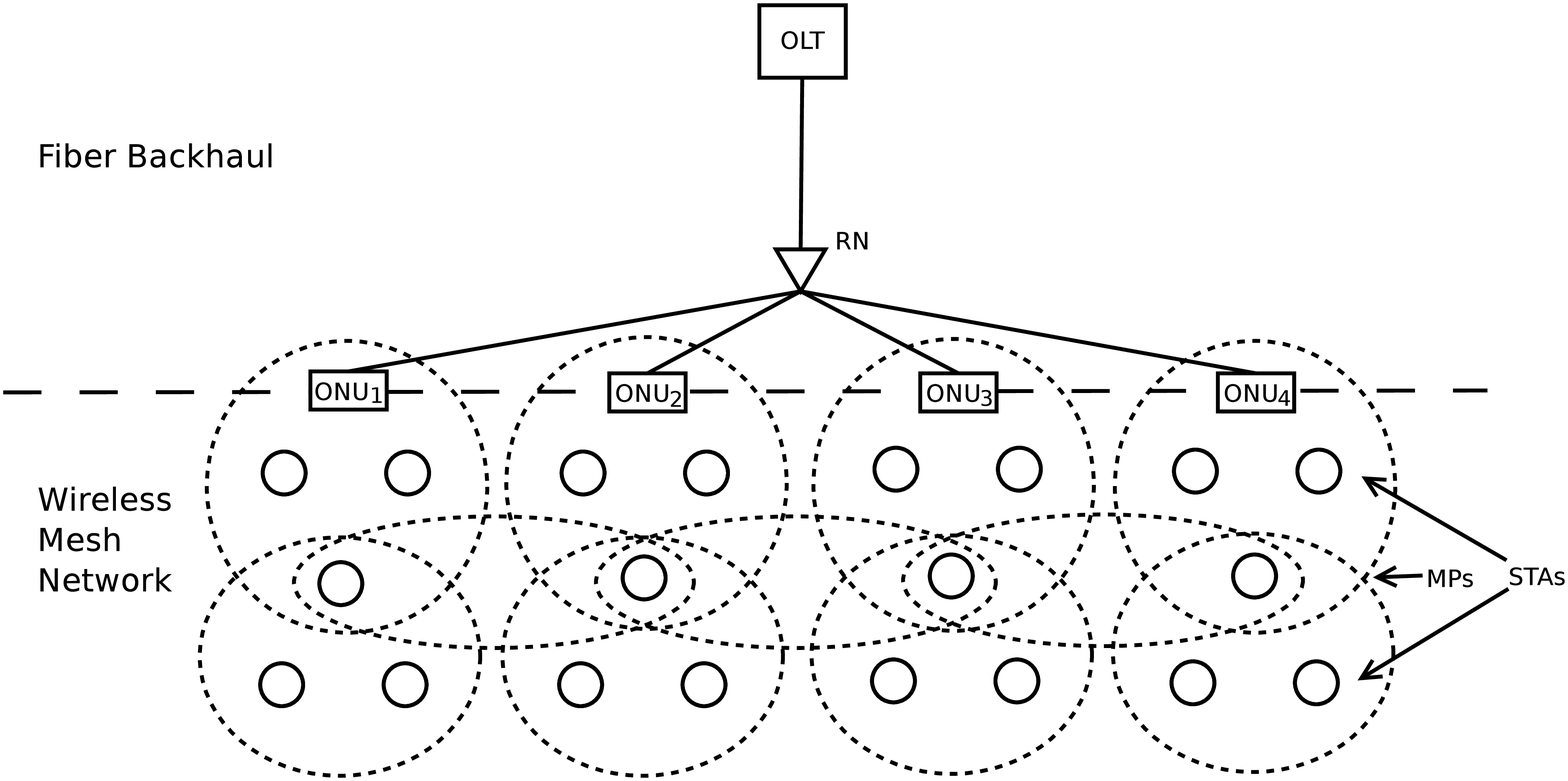}
\caption{FiWi network configuration for verifying simulations: 
4 ONU/MPPs, 4 MPs, and 16 STAs distributed over 11 wireless zones 
(dashed circles).}
\label{fig:fig4}
\end{center}
\end{figure}

In our initial verifying simulations, we consider the FiWi network
configuration of Fig.~\ref{fig:fig4}. The fiber backhaul is a TDM PON,
or a wavelength-broadcasting/routing WDM PON with
$\Lambda=2$ bidirectional wavelength channels ($\lambda=1$,
$\lambda=2$),
each operating at $c=c^{(\lambda)}=1$~Gb/s (compliant with IEEE 802.3ah).
In the case of the wavelength-routing (WR) WDM PON,
the two sectors are defined as: $\lambda=1$: $\{ONU_1,
ONU_2\}$ and $\lambda=2$: $\{ONU_3, ONU_4\}$. All four ONUs are
located 20 km from the OLT (translating
into a one-way propagation delay $\psi =\psi^{(\lambda)}=0.1$ ms) and
are equipped with an MPP. The WMN is composed of the aforementioned 4
MPPs plus 16 STAs and 4 MPs, which are distributed over 11 wireless
zones, as shown in Fig.~\ref{fig:fig4}. For instance, the WMN zone
containing $ONU_1$ comprises 1 MPP, 2 STAs, and 1 MP. MPPs and STAs
use a single radio, whereas MPs use 3, 4, 4, 3 radios from left to
right in Fig.~\ref{fig:fig4}. All WMN nodes apply the RTS/CTS access
mechanism.
The WMN operates at $r=300$ Mb/s (compliant with IEEE 802.11n)
with a bit error rate of $p_b=10^{-6}$.

\subsubsection{Traffic and Routing Assumptions}
We consider Poisson traffic with fixed-size frames of 1500 bytes
(octets). We use A-MSDU for frame aggregation, whereby each A-MSDU
carries the maximum permissible payload of 5 frames, see
Fig.~\ref{fig:fig3}(a).
Similar to~\cite{ZhWW09}, we consider two
operation modes:
($i$) \textit{WMN-only mode} which has no fiber backhaul in place;
and WMN nodes apply minimum wireless hop routing
($ii$) \textit{wireless-optical-wireless mode} which deploys the FiWi
network configuration of Fig.~\ref{fig:fig4}.
For both modes, we consider the minimum interference routing
algorithm~\cite{ZWW09},
which selects the path with the
minimum number of wireless hops.
We compare different routing algorithms in Section~\ref{fiwiroutalg:sec}.

\subsubsection{Verifying Simulations}
\begin{figure}[t]
\begin{center}
\includegraphics[width=.5\textwidth]{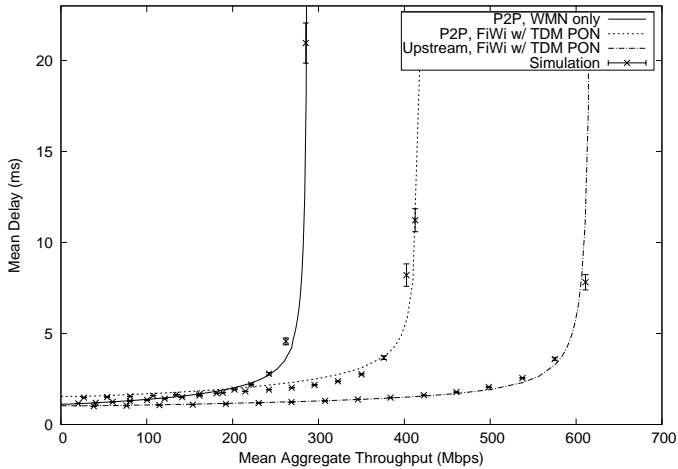}
\caption{Mean delay vs. mean aggregate throughput performance of WMN and TDM PON based FiWi networks for peer-to-peer (P2P) and upstream traffic.}
\label{fig:result1}
\end{center}
\end{figure}

The simulation results
presented in~\cite{ZhWW09} indicate that the throughput performance of
WMNs deteriorates much faster for increasing peer-to-peer traffic
among STAs than that of FiWi networks, while WMN and FiWi networks
achieve the same throughput when all traffic is destined to the
Internet. For comparison with~\cite{ZhWW09}, we consider
\textit{peer-to-peer (P2P) traffic}, where each frame generated by a
given STA is destined to any other of the remaining 15 STAs with equal
probability 1/15, and \textit{upstream traffic}, where all frames
generated by the STAs are destined to the OLT. Fig.~\ref{fig:result1}
depicts the results of our probabilistic analysis for the mean delay
as a function of the mean aggregate throughput
of a stand-alone WMN network and a TDM PON based FiWi
network for P2P and upstream traffic. The figure also shows verifying
simulation results and their 95\% confidence intervals, whereby
simulations were run 100 times for each considered traffic
load\footnote{Our simulator is based on OMNeT++ and uses the
  communication networks package \textit{inetmanet} with extensions
  for frame aggregation, wireless multihop routing, TDM/WDM PONs, and
  integrated WMN/PON routing.}.

We observe from Fig.~\ref{fig:result1}
that the mean delay of the WMN increases sharply as the mean aggregate
throughput asymptotically approaches its maximum data rate of 300
Mb/s. We also confirm the findings of~\cite{ZhWW09} that under P2P
traffic the mean aggregate throughput can be increased by using a TDM
PON as fiber backhaul to offload the wireless mesh front-end at the
expense of a slightly increased mean delay due to the introduced
upstream and downstream PON delay to and from the OLT.
As opposed to~\cite{ZhWW09}, however,
Fig.~\ref{fig:result1} shows that the throughput-delay performance of
the considered FiWi network is further improved significantly under
upstream traffic. These different observations are due to the fact
that in~\cite{ZhWW09} the single-radio single-channel WMN based on
legacy IEEE 802.11a WLAN with a limited data rate of 54 Mb/s suffered
from severe channel congestion close to the MPPs, which is
alleviated in the multi-radio multi-channel WMN based on
next-generation high-throughput WLAN technologies.

\begin{figure}[t]
\begin{center}
\includegraphics[width=.5\textwidth]{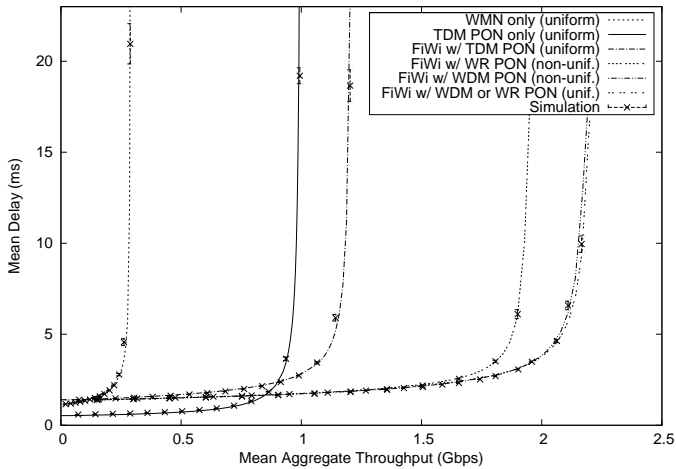}
\caption{Mean delay vs. mean aggregate throughput performance of different FiWi network architectures for uniform and nonuniform traffic.}
\label{fig:result2}
\end{center}
\end{figure}
Next, we verify different FiWi network architectures and their
constituent subnetworks for \textit{uniform} and \textit{nonuniform
  traffic} for minimum (wireless or optical) hop
routing~\cite{ZhWW09}. Fig.~\ref{fig:result2} depicts the throughput-delay
performance of a stand-alone WMN front-end, stand-alone TDM PON, and a
variety of integrated FiWi network architectures using different fiber
backhaul solutions, including conventional TDM PON,
wavelength-broadcasting WDM PON (WDM PON), and wavelength-routing WDM
PON (WR PON). In the TDM PON only (WMN only) scenario under uniform
traffic, each ONU (STA) generates the same amount of traffic and each
generated frame is destined to any of the remaining ONUs (STAs) with
equal probability. As expected, the WMN and TDM PON saturate at
roughly 300 Mb/s and 1 Gb/s, respectively, and the TDM PON is able to
support much higher data rates per source node (ONU) at lower delays
than the WMN. Furthermore, we observe from Fig.~\ref{fig:result2} that
under uniform traffic conditions, where STAs and ONUs 
send unicast traffic randomly uniformly distributed among themselves,
FiWi networks based on a wavelength-broadcasting WDM PON or a WR PON
give the same throughput-delay performance,
clearly outperforming their single-channel TDM PON based
counterpart. However, there is a clear difference between WDM PON and
WR PON fiber backhauls when traffic becomes unbalanced. To see this,
let us consider a nonuniform traffic scenario, where $ONU_1$ and
$ONU_2$ and their 4 associated STAs (see Fig.~\ref{fig:fig4}) generate
30\% more traffic than the remaining ONUs and STAs. Under such a
nonuniform traffic scenario, a FiWi network based on a
wavelength-broadcasting WDM PON performs better, as shown in
Fig.~\ref{fig:result2}. This is due to the fact that the WDM PON
provides the two heavily loaded $ONU_1$ and $ONU_2$ with access to
both wavelength channels, as opposed to the WR PON, thus resulting in
an improved throughput-delay performance.

Overall, we note that the analysis and verifying simulation results
presented in Figs.~\ref{fig:result1} and~\ref{fig:result2} match very
well for a wide range of FiWi network architectures and
traffic scenarios.

\subsection{FiWi Routing Algorithms}
\label{fiwiroutalg:sec}
Recall from Section~\ref{sec:traffic_model} that our capacity and
delay analysis flexibly accommodates any routing algorithm
and allows for multipath routing in both the fiber and wireless
domains. In this section, we study the impact of different routing
algorithms on the throughput-delay performance of next-generation FiWi
access networks in greater detail, 
including their sensitivity to key network parameters.  Specifically,
we examine the following single-path routing algorithms:

\noindent
{\bf{Minimum hop routing}:} Conventional shortest path routing selects
for each source-destination node pair the path minimizing the
required number of wireless and/or optical hops.

\noindent
{\bf{Minimum interference routing}}~\cite{ZWW09}:
The path with the minimum wireless hop count is
  selected. The rationale behind this algorithm is that the maximum
  throughput of wireless networks is typically much lower compared to
  the throughput in optical networks. Thus, minimizing the
  wireless hop count tends to increase the maximum FiWi network
   throughput.

\noindent
{\bf{Minimum delay routing}:} Similar to the previously
  proposed WMN routing algorithms DARA~\cite{SYDM08},
  CaDAR~\cite{RRSG09}, and RADAR~\cite{SYDM07}, we apply a slightly
  extended minimum delay routing algorithm, which aims at selecting
  the path that minimizes the end-to-end delay of Eq.~(\ref{eq:delay})
  across the entire bimodal FiWi access network.
  The applied minimum delay routing algorithm is a greedy algorithm
  and proceeds in two steps. In the initialization step, paths are set
  to the minimum hop routes. The second
  step computes for each source-destination node pair
  the path with the minimum end-to-end delay under given
  traffic demands.

\noindent
{\bf{Optimized FiWi routing algorithm (OFRA)}:} We propose the
optimized FiWi routing algorithm (OFRA), which
proceeds in two steps similar to minimum delay routing.
After the initialization step to minimum hop routes, the second step
of OFRA computes for each source-destination node pair the
path $p$ with the minimization objective
\begin{equation}
\label{eq:OFRA}
\min_{p}\left (\sum_{\forall n \in p} (\rho_n) + \max_{\forall n \in p}(\rho_n)\right ),
\end{equation}
where $\rho_n$ represents the long-run 
traffic intensity at a generic FiWi
network node $n$, which may be either an optical node belonging to the
fiber backhaul or a wireless node belonging to the wireless mesh
front-end.
Based on a combination of historic traffic patterns as well as
traffic measurements and estimations 
similar to~\cite{BiFG05,ElMSW03,GeMu03}, 
the traffic intensities $\rho_n$ used in
OFRA can be periodically updated with strategies
similar to~\cite{RRSG09,FlJa94}.
These long-run traffic intensities vary typically slowly,
e.g., with a diurnal pattern, allowing essentially offline computation
of the OFRA paths.
More precisely, for the WR PON we have
$\rho_n=\rho^{d,\lambda}$ (see Eq. (\ref{eq:WR_OLT})) if node $n$ is
the OLT and $\rho_n=\rho^{u,\lambda}$ (see Eq. (\ref{eq:WR_ONU})) if
node $n$ is an ONU. For the wavelength-broadcasting TDM and WDM PON,
we have $\rho_n=\rho^d$ (Eq. (\ref{eq:rho_d})) and $\rho_n=\rho^u$
(Eq. (\ref{eq:rho_u})) if node $n$ is the OLT or an ONU,
respectively.
For a wireless node $n$, $\rho_n$ is given by the left-hand side 
of~(\ref{eq:rho_omega}).

OFRA's path length measure includes
the maximum traffic intensity $\max_{\forall n \in p}(\rho_n)$ along a path $p$
in order to penalize paths with a high traffic intensity
at one or more FiWi network nodes.
For a given set of traffic flows, OFRA minimizes the
  traffic intensities, particularly the high ones, at the FiWi network nodes.
  Decreasing the
  traffic intensities tends to allow for a higher number of
 supported traffic flows and thus higher throughput.

To allow for a larger number of possible paths for the following
 numerical investigations of the different considered routing algorithms,
we double the FiWi network configuration of Fig.~\ref{fig:fig4}.
We consider a wavelength-routing (WR) WDM PON with
a total of 8 ONU/MPPs, 8 MPs, and 32 STAs in 22 wireless zones, whereby
ONU/MPPs 1-4 and ONU/MPPs 5-8 are served on wavelength channel
$\lambda=1$ and $\lambda=2$, respectively. Furthermore, to evaluate
different traffic loads in the optical and wireless domains, we consider
the following traffic matrix for the OLT, $O$ ONUs, and $N$
STAs:\\ \\ \bordermatrix{~ & 0 & 1 & \dots & O & O+1 & \dots & O + N
  \cr 0 & 0 & B\alpha & B\alpha & B\alpha & \alpha & \alpha & \alpha
  \cr 1 & B\alpha & 0 & B\alpha & B\alpha & \alpha & \alpha & \alpha
  \cr \vdots & B\alpha & B\alpha & 0 & B\alpha & \alpha & \alpha &
  \alpha \cr O & B\alpha & B\alpha & B\alpha & 0 & \alpha & \alpha &
  \alpha \cr O+1 & \alpha & \alpha & \alpha & \alpha & 0 & \alpha &
  \alpha \cr \vdots & \alpha & \alpha & \alpha & \alpha & \alpha & 0 &
  \alpha\cr O+N & \alpha & \alpha & \alpha & \alpha & \alpha & \alpha
  & 0 \cr}, \\ \\ \\
where $\alpha\geq0$ denotes the mean traffic rate (in frames/second).
The parameter $B\geq1$ can be used to test different traffic
intensities in the PON, since
  the ONUs could be underutilized compared to the WMN in the
  considered topology.
Recall from Fig.~\ref{fig:fig1} that ONUs may
serve multiple subscribers with wired ONU access,
whose aggregate traffic leads to an increased load at ONUs.

\begin{figure}[t]
\begin{center}
\includegraphics[width=.5\textwidth]{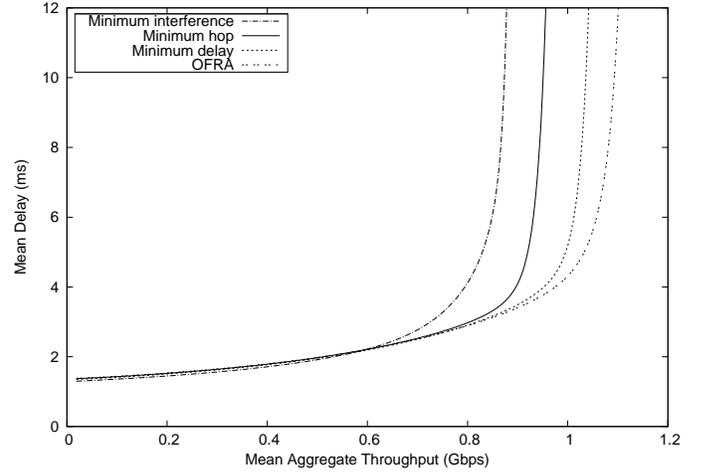}
\caption{Mean delay vs. mean aggregate throughput performance of
  different FiWi routing algorithms for a conventional
  wavelength-routing (WR) WDM PON of 20 km range and $B=1$.}
\label{fig:result3}
\end{center}
\end{figure}
For a conventional WR WDM PON with a typical optical
fiber range of 20 km, Fig.~\ref{fig:result3} illustrates that OFRA
yields the best throughput-delay performance for $B=1$, i.e., every
optical and wireless FiWi node generates the same amount of
traffic.
Minimum interference routing tends to overload the wireless
MPP interfaces as it does not count the fiber backhaul as a hop,
resulting in high delays.

\begin{figure}[t]
\begin{center}
\includegraphics[width=.5\textwidth]{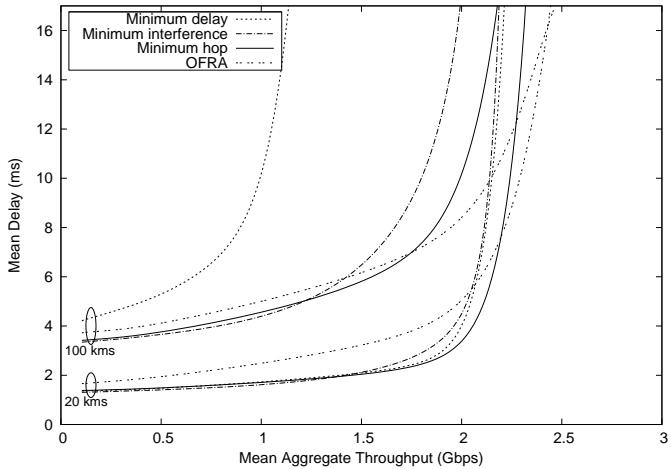}
\caption{Mean delay vs. mean aggregate throughput performance 
of different FiWi routing algorithms for 
($i$) a conventional 20 km range and 
($ii$) a 100 km long-reach wavelength-routing (WR) WDM PON and $B=100$.}
\label{fig:result4}
\end{center}
\end{figure}

The throughput-delay performance of the four
considered FiWi routing algorithms largely depends on the given
traffic loads and length of the fiber backhaul.
Fig.~\ref{fig:result4} depicts their throughput-delay performance for
($i$) a conventional 20 km range and ($ii$) a 100 km long-reach
WR WDM PON, whereby in both configurations we set
$B=100$, i.e., the amount of generated traffic among optical nodes
(OLT and ONUs) is 100 times higher than that between
node pairs involving at least one wireless node (STA). More precisely,
all the (increased) inter-ONU/OLT traffic is sent
across the WDM PON only, thus creating a higher loaded fiber
backhaul. We observe from Fig.~\ref{fig:result4} that in general all
four routing algorithms achieve a higher maximum
aggregate throughput due to the increased traffic load carried on the
fiber backhaul.

We observe that for a conventional
20 km range WR WDM PON with small to medium traffic loads,
OFRA gives slightly higher delays than the other
considered routing algorithms.
This observation is in contrast to Fig.~\ref{fig:result3},
though in both figures
OFRA yields the highest maximum aggregate throughput.
We have measured the
traffic on the optical and wireless network interfaces of each
ONU/MPP. Our measurements show that at low to medium traffic loads
with $B = 100$,
OFRA routes significantly less traffic across the WDM PON than the
other routing algorithms, but
instead uses the less loaded wireless mesh front-end. This is due to
the objective of OFRA to give preference to links with lower traffic
intensities. As a consequence, for $B=100$, OFRA routes relatively
more traffic over
lightly loaded wireless links, even though this implies more
wireless hops, resulting in a slightly increased mean delay compared to
the other routing algorithms at low to medium loads.

Fig.~\ref{fig:result4} also shows the impact of the increased
propagation delay in a long-reach WDM PON with a
fiber range of 100 km between OLT and ONUs.
Aside from a generally
increased mean delay, we observe that minimum hop
and minimum interference routing as well as OFRA
provide comparable delays at low to medium traffic loads, while the
maximum achievable throughput differences at high traffic loads
are more pronounced than for the 20~km range.
The favorable performance of
OFRA at high traffic loads is potentially of
high practical relevance since access networks are the bottlenecks
in many networking scenarios and thus experience relatively high
loads while core networks operate at low to medium loads.

Fig.~\ref{fig:result4} illustrates that
minimum delay routing performs poorly in long-reach WDM PON based
FiWi access networks. Our measurements indicate that minimum
delay routing utilizes the huge bandwidth of the long-reach
WDM PON much less than the other routing algorithms in order to
avoid the increased propagation delay. As a consequence, with minimum
delay routing most traffic is sent across the WMN, which offers
significantly lower data rates than the fiber backhaul, resulting in a
congested wireless front-end and thereby an inferior throughput-delay
performance. 

\subsection{Fiber Failures}
To highlight the flexibility of our analysis, we
note that it accommodates any type and number of fiber
failures.
Fiber failures represent one of the major differences
between optical (wired) fiber and wireless networks and affect the
availability of bimodal FiWi networks. 
In the event of one or more
distribution fiber cuts, the corresponding disconnected ONU/MPP(s)
turn(s) into a conventional wireless MP without offering gateway
functions to the fiber backhaul any longer.
FiWi access network survivability for arbitrary fiber failure
  probabilities has been analyzed in~\cite{GhSM11}.

\begin{figure}[t]
\begin{center}
\includegraphics[width=.5\textwidth]{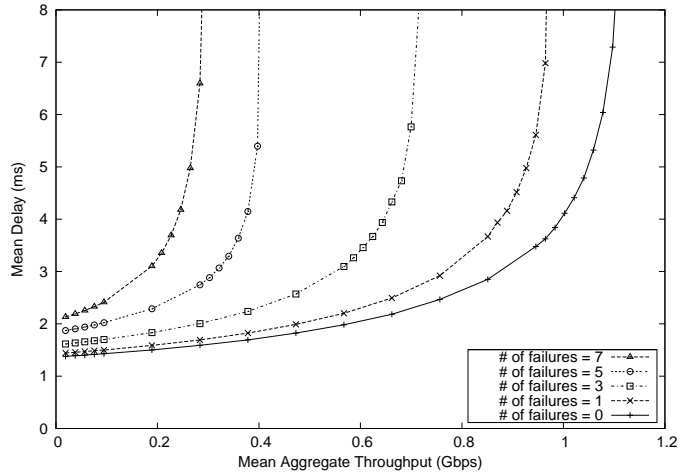}
\caption{Impact of distribution fiber failures on throughput-delay
  performance of a 20 km range wavelength-routing WDM PON using OFRA
  with $B=1$.}
\label{fig:result5}
\end{center}
\end{figure}

Fig.~\ref{fig:result5} illustrates the detrimental impact of distribution
fiber failures on the throughput-delay performance of a 20 km range
wavelength-routing WDM PON, which is typically left unprotected due to
the small number of cost-sharing subscribers and cost-sensitivity of
access networks.
We also note that the analytical framework
  is able to account for other types of network failure, e.g., ONU/MPP
  failures. In this case, malfunctioning ONU/MPPs become unavailable
  for both optical and wireless routing.

In principle, FiWi access networks can be made
  more robust against fiber failures through various optical
  redundancy strategies, such as ONU dual homing, point-to-point
  interconnection fibers between pairs of ONUs, fiber protection rings
  to interconnect a group of closely located ONUs by a short-distance
  fiber ring, or meshed PON topologies~\cite{Maier12}.
These redundancy strategies in general imply
  major architectural and ONU modifications of the FiWi access network
  of Fig.~\ref{fig:fig1} under consideration. To incorporate such
  topological PON modifications, the fiber part of the capacity and
  delay analysis would need to be modified accordingly.

\subsection{Very High Throughput WLAN}
Our analysis is also applicable to the emerging 
IEEE standard 802.11ac for future VHT WLANs with
raw data rates up to 6900 Mb/s.
In addition to a number of PHY layer enhancements,
IEEE 802.11ac will increase the maximum A-MSDU size from 7935 to 11406
octets and the maximum A-MPDU size from 65535 octets to 1048575
octets. Both enhancements can be readily accommodated in our
analysis by setting the parameters $A_{\max}^{\text{A-MSDU/A-MPDU}}$ and
$r$ accordingly.

\begin{figure}[t]
\begin{center}
\includegraphics[width=.5\textwidth]{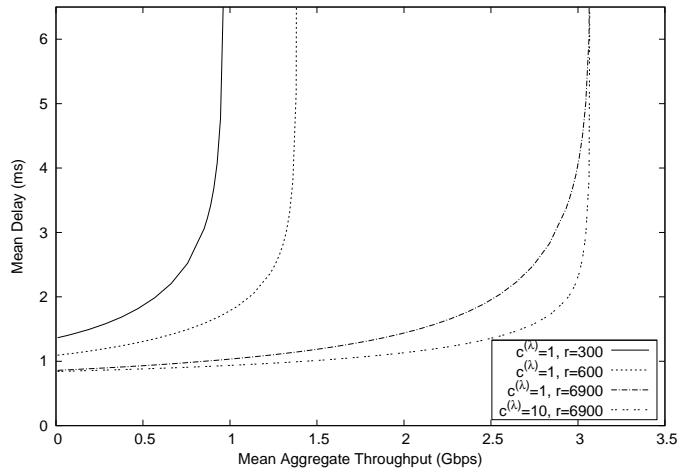}
\caption{Throughput-delay performance of next-generation FiWi access
  networks based on high-speed wavelength-routing WDM PON and VHT WLAN
  technologies using minimum hop routing with $B=1$ (optical and
  wireless data rates , $c^{(\lambda)}$ and $r$, are given in Gb/s and
  Mb/s, respectively).}
\label{fig:result6}
\end{center}
\end{figure}
Fig.~\ref{fig:result6} illustrates the FiWi network
performance gain achieved with a wireless front-end based on VHT
WLAN instead of IEEE 802.11n WLAN with maximum data
rate of 600 Mb/s, for minimum hop routing, an optical range of 20~km,
and $B=1$. For a wavelength-routing WDM PON operating at a
wavelength channel data rate of 1 Gb/s, we observe from
Fig.~\ref{fig:result6} that VHT WLAN roughly triples the maximum mean
aggregate throughput and clearly outperforms 600 Mb/s 802.11n WLAN in
terms of both throughput and delay. Furthermore, the figure shows that
replacing the 1 Gb/s wavelength-routing WDM PON with its
high-speed 10~Gb/s counterpart
(compliant with the IEEE 802.3av 10G-EPON standard)
does not yield a higher maximum
aggregate throughput, but it does lower the mean delay especially
at medium traffic loads before wireless links at the optical-wireless
interfaces get increasingly congested at higher traffic loads.

\section{Conclusions}
\label{sec:conclusions}
A variety of routing algorithms have recently been proposed for
integrated FiWi access networks based on complementary EPON and
WLAN-mesh networks.
In this article, we presented the first analytical framework to
quantify the performance of FiWi network routing algorithms, validate
previous simulation studies, and provide insightful guidelines for the
design of novel integrated optical-wireless routing algorithms for
future FiWi access networks leveraging next-generation PONs,
notably long-reach 10+ Gb/s TDM/WDM PONs, and emerging Gigabit-class
WLAN technologies. Our analytical framework is very flexible and can
be applied to any existing or new optical-wireless routing
algorithm. Furthermore, it takes the different characteristics of
disparate optical and wireless networking technologies into
account. Beside their capacity mismatch and bit error rate
differences, the framework also incorporates arbitrary frame size
distributions, traffic matrices, optical/wireless propagation delays,
data rates, and fiber cuts.
We investigated the performance of minimum
hop, minimum interference (wireless hop), minimum delay,
and our proposed OFRA routing algorithms. The
obtained results showed that OFRA yields the highest maximum aggregate
throughput for both conventional and long-reach wavelength-routing WDM
PONs under balanced and unbalanced traffic loads. For a higher loaded
fiber backhaul, however, OFRA gives priority to lightly loaded
wireless links, leading to an increased mean delay at small to medium
wireless traffic loads. We also observed that using VHT WLAN helps
increase the maximum mean aggregate throughput significantly, while
high-speed 10 Gb/s WDM PON helps lower the mean delay especially at
medium traffic loads.

There are several important directions for future research.
One direction is to examine mechanisms for providing
qulity of service or supporting specific traffic types,
see e.g.,~\cite{DiLD13,WeAMR13}.
Further detailed study of the impact of different dynamic bandwidth
allocation approaches for long-reach PONs, 
e.g.,~\cite{AhCW13,BuAT13,KiFSM13,JiMFD12,MeMR13,MeJDF10}, 
and their effectiveness in integrated FiWi networks is of interest.
Yet another direction is the examine the internetworking
of FiWi networks with metropolitan area 
networks~\cite{BiBC13,MaRW03,ScMRW03,YaMRC03,YuCL10,WRSK03}.

\bibliographystyle{ieeetran}


\end{document}